\documentclass{article}

     \PassOptionsToPackage{numbers, compress}{natbib}

     \usepackage[preprint]{neurips_2020}

\usepackage{natbib}
\usepackage[utf8]{inputenc} 
\usepackage[T1]{fontenc}    
\usepackage{hyperref}       
\usepackage{url}            
\usepackage{booktabs}       
\usepackage{amsfonts}       
\usepackage{nicefrac}       
\usepackage{microtype}      
\usepackage{graphicx}       
\usepackage{multirow}       
\usepackage{multicol}       
\usepackage{amsmath}       
\usepackage{xcolor} 
\usepackage{subfig}         

\usepackage{graphicx}
\usepackage{subfig}

\usepackage{latexsym,amsmath,amsfonts,amssymb,amsthm}
\usepackage{setspace,multirow,color,comment,array}
\usepackage{url}
\usepackage{enumerate} 
\usepackage{arydshln}
\usepackage{breakcites}
\usepackage{pifont}
\usepackage{lipsum} 

\usepackage{times}
\usepackage[normalem]{ulem}

\usepackage{multirow} 
\usepackage{wrapfig}
\usepackage{textcomp}  
 
\usepackage{algorithm,algpseudocode}
\usepackage{algorithmicx} 
\algnewcommand\algorithmicinput{\textbf{Input:}}
\algnewcommand\algorithmicoutput{\textbf{Output:}}
\algnewcommand\Input{\item[\algorithmicinput]}%
\algnewcommand\Output{\item[\algorithmicoutput]}%

\algnewcommand\algorithmicswitch{\textbf{switch}}
\algnewcommand\algorithmiccase{\textbf{case}}
 
\errorcontextlines\maxdimen

\makeatletter

\usepackage{amsthm}

\usepackage{xcolor}
\definecolor{purple}{HTML}{9F4C7C}

\newtheorem{Thm}{Theorem} 
\newtheorem{Def}{Definition}

\newtheorem{Remark}{Remark}

\newcommand{\ie}{{{i.e.,~}}}
\newcommand{\eg}{{{e.g.~}}}
\newcommand{\etal}{{{et~al.~}}}

\newcommand{\niter}{T}
\newcommand{\ep}{{\epsilon}}
\newcommand{\la}{{\leftarrow}}

\newcommand{\bS}{\boldsymbol S}
\newcommand{\bs}{\boldsymbol s}

\newcommand{\bV}{\boldsymbol V}
\newcommand{\bx}{\boldsymbol x}
\newcommand{\bX}{\boldsymbol X}
\newcommand{\bz}{\boldsymbol z}
\newcommand{\bbeta}{\boldsymbol \beta}
\newcommand{\btheta}{\boldsymbol \theta}
\newcommand{\bmu}{\boldsymbol \mu}

\newcommand{\bbN}{{\mathbb N}}

\newcommand{\bbR}{{\mathbb R}}
\newcommand{\bbZ}{{\mathbb Z}}

\newcommand{\D}{{\mathcal D}}
\newcommand{\N}{{\mathcal N}}

\newcommand{\sfm}{\mathsf m}
\newcommand{\sfM}{\mathsf M}

\newcommand{\bI}{\textbf I}

\newcommand{\mmax}{\mathsf{max}}

\newcommand{\diag}{\mathsf{diag}}

\newcommand{\SGLD}{\textsc{SGLD}}

\newcommand{\Ag}{\mathsf{Ag}}

\newcommand{\pk}{\mathsf{pk}}
\newcommand{\sk}{\mathsf{sk}}

\newcommand{\evk}{\mathsf{evk}}

\newcommand{\Enc}{\mathsf{Enc}}

\newcommand{\Rot}{\mathsf{Rot}}
\newcommand{\Dec}{\mathsf{Dec}}

\newcommand{\ct}{\mathsf{ct}}
\newcommand{\HM}{\mathsf{HM}}
\newcommand{\SM}{\mathsf{SM}}

\newcommand{\ceil}[1]{\lceil{#1}\rceil}

\newcommand{\rd}[1]{\lfloor{#1}\rceil}

\newcommand{\RNum}[1]{\uppercase\expandafter{\romannumeral #1\relax}}

\newcommand{\RN}[1]{\textup{\uppercase\expandafter{\romannumeral#1}}}

\newif\ifdraft
\drafttrue
\ifdraft
\newcommand{\todo}[1]{ {\textcolor{red} { ***TODO: #1 }}}
\else
\newcommand{\todo}[1]{}
\fi

\title{Secure and Differentially Private Bayesian Learning  on Distributed Data}

\author{%
  Yeongjae Gil\\
  School of Management Engineering,\\
  Ulsan National Institute of \\
  Science and Technology\\
  \texttt{yjgil123@unist.ac.kr} \\
  \And
  Xiaoqian Jiang\\
  School of Biomedical Informatics,\\ 
  University of Texas,\\
  Health Science Center at Houston\\
  \texttt{Xiaoqian.Jiang@uth.tmc.edu} \\
  \And
  Miran Kim\\
  School of Biomedical Informatics,\\ 
  University of Texas,\\
  Health Science Center at Houston\\
  \texttt{Miran.Kim@uth.tmc.edu} \\
  \And
  Junghye Lee\\
  School of Management Engineering,\\
  Ulsan National Institute of \\
  Science and Technology\\
  \texttt{junghyelee@unist.ac.kr} \\
}

\begin{document}

\maketitle


\begin{abstract}
Data integration and sharing maximally enhance the potential for novel and meaningful discoveries. However, it is a non-trivial task as integrating data from multiple sources can put sensitive information of study participants at risk. To address the privacy concern, we present a distributed Bayesian learning approach via Preconditioned Stochastic Gradient Langevin Dynamics with RMSprop, which combines differential privacy and homomorphic encryption in a harmonious manner while protecting private information. We applied the proposed secure and privacy-preserving distributed Bayesian learning approach to logistic regression and survival analysis on distributed data, and demonstrated its feasibility in terms of prediction accuracy and time complexity, compared to the centralized approach. 
\end{abstract}

\section{Introduction}\label{intro}

In the past few decades, machine learning has shown significant success in a wide variety of real-world applications. 
In particular, distributed learning has shown its potential to effectively leverage decentralized data to collaboratively train machine learning models.  
However, it might put sensitive information at risk as the training process involves intermediary parameter exchange, which might be used by adversaries to conduct inference attacks to reconstruct training samples or infer the membership~\cite{F+14, S+17}.

Several cryptography technologies have been recently developed to support privacy-preserving computation on distributed data. 
Among these solutions, differential privacy (DP) is a cryptography-motivated privacy protection mechanism to protect private information through random perturbation in general learning procedures such as machine learning algorithms~\cite{Dwork06, Dwork2006-sx}. However, its strictness to guarantee a strong privacy applies to the output of the algorithm, which might introduce too much noise in cases where the transfer or exchange of intermediate statistics (including the output), such as distributed learning, is required.
Recently, homomorphic encryption (HE) has shed light on privacy-preserving distributed learning challenges, which allows encrypted data to be computed without having access to raw data~\cite{Gen09}. 
However, HE is computationally intensive, especially when the evaluation of a deep circuit is needed {since multiplication operations bring about increased noise level or decreased ciphertext modulus}. 
When HE is combined with DP, it generates a positive synergistic effect: (1) reduced DP noise due to the protection of intermediary statistics and (2) minimized HE computation thanks to the privacy protection of DP.
Although such studies have been successfully conducted on generalized linear models using a combination of DP and HE~\cite{A+16, KLOJ19}, this framework does not directly apply to Bayesian ones. 

Bayesian models have become recognized as one of the powerful machine learning techniques due to their ability to leverage prior information provided by domain experts and statistical evidence from data (i.e., confidence estimation) and to avoid overfitting. Despite the advantages of Bayesian methods, they have one important disadvantage - high computational complexity. 
One remarkable recent approach to resolve the computational challenges of Bayesian learning is to combine stochastic gradient descent (SGD)~\cite{RM51} and Monte Chain Monte Carlo (MCMC) methods~\cite{N95}. 
When an appropriate sequence of step sizes is given, it has been shown to converge to the true posterior based on the gradient on a minibatch of data in several variants~\cite{A+12, C+14, D+14}. In this paper, we will focus on the pioneering and the most common algorithm, Stochastic Gradient Langevin Dynamics (SGLD)~\cite{WT11}. More specifically, the preconditioned version of SGLD by RMSprop, developed to increase the efficiency of the standard SGLD~\cite{WT11, Li+16}, is our interest. 
There has also been work on differentially private SGLD~\cite{W+15, Li+19}, but no work for the preconditioned SGLD. 

There are other approaches for privacy-preserving distributed Bayesian learning which combine DP and multi-party computation (MPC).
Heikkil{\"a} et al.~\cite{H+17} proposed the hybrid system that allows data owners to secretly share data with several independent servers and the servers to securely compute the required sums of sufficient statistics of Bayesian learning.
So, if taking into account Bayesian learning which needs different types of computation (\eg multiplication), their approach may require additional computational cost and communication between servers.


In this paper, we introduce a general framework that combines the strength of DP and HE in order to support adaptive and efficient optimizers for Bayesian learning. 
Our approach not only protects the training data during the learning process using HE, but also preserves the privacy of the resulting model via DP. 
We show that under standard assumptions, the Preconditioned SGLD (pSGLD) algorithm with the RMSprop preconditioner is differentially private. 
We present an efficient method to securely train Bayesian learning models in a distributed setting. 
We make use of the diagonal preconditioner to exploit parallel computation over packed ciphertexts.
As a result, it requires one HE computation protocol execution per iteration but with a trivial per-iteration overhead in terms of computation and communication.
Finally, we conduct extensive empirical evaluation to show its scalability on real data. 
We apply the proposed secure and privacy-preserving distributed Bayesian learning approach to logistic regression and survival analysis on distributed data, and demonstrate its feasibility in terms of prediction accuracy and time complexity, compared to the centralized approach.  

The results have demonstrated that our approach (when applied on distributed data) can produce secure and privacy-preserving predictive models with: (1) good accuracy comparable to the global model trained on centralized data and (2) high-efficiency thanks to the synergistic combination of DP and HE.
{Our novel method enjoys ``privacy for free in a distributed scenario" because of the HE protection of intermediary statistics, while the theoretical support of DP's privacy guarantee on parameters allows the construction of a very shallow circuit even for the iterative learning algorithm.  
Consequently, it enables us to avoid an expensive bootstrapping procedure that refreshes low-level ciphertexts (\eg Cheon \etal\cite{CHK+18a}), and so our proposed approach is computationally tractable compared to HE-based solutions.}

\section{Preliminaries}


\subsection{Differential Privacy}\label{ssec:DP}
DP is one of the strongest privacy-preserving criteria to provide quantitative privacy guarantees based on a probabilistic formulation without any assumption about adversary. 

\begin{Def}[($\epsilon,\delta$)-Differential Privacy~\cite{Dwork06}]
    {Given a pair of data sets $\mathcal{D}$ and $\mathcal{D'}$ differing by at most one record (\ie $|\mathcal{D}\Delta\mathcal{D'}|\leq 1$), a randomized algorithm $\Ag$ satisfies ($\epsilon,\delta$)-DP if and for all $\mathcal{S} \subseteq Range(\Ag)$, we have 
    $ {\Pr[\Ag(\mathcal{D})\in \mathcal{S}]\leq e^{\epsilon} \cdot \Pr[\Ag(\mathcal{D'})\in \mathcal{S}]+\delta,}$
    where the probabilities are over the randomness of the algorithm $\Ag$. 
    The privacy parameters $\epsilon$ and $\delta$ are positive numbers, for which small numbers correspond to stronger protection and vice versa. When $\delta=0$, $\Ag$ is $\epsilon$-differentially private}. 
\end{Def}

Any adversaries cannot distinguish the output from $\mathcal{D}$ or $\mathcal{D'}$ as long as $\epsilon$ and $\delta$ are small enough. 
Thus, any record that makes the difference between $\mathcal{D}$ and $\mathcal{D'}$ is protected, and therefore, providing an ad omnia protection to the entire dataset.
Adding controlled noise from predetermined distributions is a way of designing DP mechanisms. Commonly used distributions are Laplace and Gaussian distributions. With the Gaussian mechanism that adds Gaussian noise, the noise is usually calibrated to the $\ell_2$ sensitivity.

\begin{Def}[$\ell_2$~Sensitivity~\cite{Dwork06}]
    {For any function $f:\D\rightarrow \mathbb{R}^{d}$, the sensitivity of $f$ is defined as $\Delta_2 f = \sup_{D,D'} \lVert f(\mathcal{D})-f(\mathcal{D'})\lVert_2$, for all $\mathcal{D}$ and $\mathcal{D'}$.} 
\end{Def} 

\smallskip
\begin{Thm}
 [Gaussian Mechanism~\cite{DR14}]
 Given any function $f:\D\rightarrow \bbR^d$, the Gaussian Mechanism is defined as $\hat{f}(X)=f(X)+\mathcal{N}(0,\sigma^2 {\textbf{\bI}})$. 
 For arbitrary $\epsilon,\delta\in(0,1)$, the  Gaussian Mechanism is $(\epsilon,\delta)$-differential private if $\sigma \ge \Delta_2 f(\sqrt{2\log(1.25/\delta)}/\epsilon)$.

\end{Thm}


DP also has a favorable property such as composability. 
The composition of $k$ differentially private mechanisms, where the $i$-th mechanism is ($\epsilon_i, \delta_i$)-differentially private, for $1\leq i \leq k$, is ($\sum_i \epsilon_i, \sum_i \delta_i$)-differentially private. Better results for ($\epsilon, \delta$)-differential privacy will follow from using the advanced composition theorem.

\smallskip
\begin{Thm}[Advanced Composition~\cite{DRV10}]
For every  $\epsilon > 0,\delta,\delta'>0,k\in\bbN$, and ($\epsilon,\delta$)-differentially private algorithms $\Ag_1,\Ag_2,...,\Ag_k$, the composition \textnormal{(}$\Ag_1,\Ag_2,...,\Ag_k$\textnormal{)} satisfies ($\epsilon_\textrm{g},k\delta+\delta'$)-differential privacy for $\epsilon_\textrm{g}=\sqrt{2k\log(1/\delta')}\epsilon+k\epsilon(e^\epsilon-1)$.
\end{Thm} 


\subsection{Homomorphic Encryption}\label{ssec:HE}

HE allows one to perform arithmetic operations on encrypted data and receive an encrypted result corresponding to the result of operations performed in plaintext. 
This technology has great potential in many real-world applications such as statistical testing, neural networks, and other machine learning models~\cite{LNV11, Cryptonets, JKLS18, SHE, BGR19}. 
Among the state-of-the-art HE cryptosystems, the CKKS scheme~\cite{CKKS17} is capable of performing approximate computation on encrypted data, so it has shown remarkable performance advantages in real-world applications that do not require absolute precision, which is the case for many machine learning models~\cite{KSW+18,KSK+18,JKLS18}.

A key feature of the CKKS scheme is to use a built-in rescaling operation on encrypted data as if rounding off significant digits in plain fixed-point computation. This technique leads to precision adjustment to get rid of accumulated extra digits after homomorphic computation and therefore enables to control the magnitude (size) of messages. 
We multiply a scale factor of $\Delta$ to plaintexts and convert them into the nearest integers before encryption in order to minimize the accuracy loss during computation. It is a common practice to perform the rescaling procedure by a factor of $\Delta$ on ciphertexts after each multiplication to maintain the precision of the plaintext.
Also, one can encrypt multiple plaintext values into a single packed ciphertext to perform parallel homomorphic operations in a single instruction multiple data~(SIMD) manner.
To be precise, the CKKS scheme provides an encryption function $\Enc(\cdot)$ and a decryption function $\Dec(\cdot)$ such that for $x,y \in \mathbb{R}^{n/2}$, 
$$\Dec(\Enc(x) + \Enc(y)) \approx x \oplus  y, \quad
\Dec(\Enc(x) \cdot \Enc(y)) \approx x \odot y,$$
where $n$ is defined as a power-of-two integer for an underlying cyclotomic ring dimension, and $\oplus$ and $\odot$ denote the element-wise addition and multiplication over real numbers.

\subsection{Preconditioned Stochastic Gradient Langevin Dynamics}
The vanilla SGLD is composed of characteristics from SGD and Langevin dynamics, a mathematical extension of molecular dynamics models. Given data $\mathcal{D}= \{ \bx_i \in\mathbb{R}^p\} _{i=1}^N,$ 
the posterior of model parameters $\btheta$ with prior $p(\btheta)$ and likelihood $\prod_{i=1}^N p(\bx_i|\btheta)$ is computed as $p(\btheta|\D) \propto p(\btheta) \prod_{i=1}^N p(\bx_i|\btheta)$. 
The parameter update is computed as:
$${\Delta\btheta_{t}= -\frac{\eta_t}{2} \left( \nabla_{\btheta} \log{p (\btheta_t)}+\frac{N}{\tau}\sum_{i=1}^{\tau}\nabla_{\btheta} \log {p(\bx_{t_i}|\btheta_t)}\right)+ \bz_t,}
$$
\noindent where $\eta_t$ is a sequence of step sizes and $\bz_t \sim \mathcal{N}(0,\eta_t \bI)$ with $\bI$ denoting the identity matrix. This update function is based on a perturbed version of the negative log-posterior objective function on the SGD with a mini-batch size $\tau$. 
SGLD incorporates uncertainty into estimates to avoid converging to the maximum a posterior point estimate. 


To increase the efficiency of the standard \SGLD~\cite{WT11}, which updates with the same step size, an improved solution is to employ preconditioning $G(\btheta)$ that aims to constitute a local transform such that the rate of curvature of the gradients is equal in all directions:
$${
\btheta_{t+1}  = \btheta_t- \frac{\eta_t}{2} \biggl[G(\btheta_t)\biggl(\nabla_\theta\log p (\btheta_t)}  {\quad + \frac{N}{\tau}\sum_{i=1}^{\tau}\nabla_\theta\log p(\bx_{t_i}|\btheta_t)\biggr)+\Gamma(\btheta_t)\biggr] +\bz_t,
}$$
where $\boldsymbol{z}_t \sim \mathcal{N}(0,\eta_t G(\btheta_t))$ and $\Gamma(\btheta)=\sum_j \frac{\partial G_{i,j}(\btheta)}{\partial \theta_j}$. 
Among many preconditioners, Adagrad~\cite{J+11}, Adam~\cite{KB15}, and RMSprop~\cite{TH12} are popular choices due to their simplicity and wide applicability.

\section{Proposed Approach}



\subsection{Privacy Analysis for Preconditioned Stochastic Gradient Langevin Dynamics with RMSprop}

This study specifies the use of RMSprop preconditioner:
{{
\begin{align*}
 V(\btheta_t) &= \alpha V(\btheta_{t-1})+(1-\alpha)\bar{g}(\btheta_t;\D_t)\odot\bar{g}(\btheta_t;D_t), \\
 G(\btheta_t) &= \diag(\textbf{1}\oslash (\lambda\textbf{1}+\sqrt{V(\btheta_t)}),
\end{align*}}}
where $\alpha\in[0,1]$ is a parameter that adjusts the balance between historical and current gradients and $\lambda\in\mathbb{R}^+$ is a curvature controlling parameter which is usually a small value. 
Here, $\odot$ and $\oslash$ operators represent element-wise matrix product and division, respectively. Since this preconditioner is updated sequentially using only the current gradient information, and only estimates a diagonal matrix, we can efficiently incorporate it with HE by utilizing SIMD based operations. In fact, Li \etal\cite{Li+16} has shown the finite-time convergence properties of pSGLD using the RMSprop preconditioner.

This section gives differential privacy analysis of pSGLD with RMSprop. We note that the pioneering and recognized work~\cite{W+15} already gave similar theoretical results on SGLD and we extend the theory to consider the use of preconditioner $G(\btheta)$. Based on the Gaussian mechanism, they showed that SGLD is $(\ep,\delta)$-differentially private for free if the step size $\eta_t$ is chosen appropriately (\ie $\eta_t < \frac{\epsilon^2 N}{128L^2\log(2.5T/\delta)\log(2/\delta)t}$ in their study). 

\medskip
\begin{Thm}
 [Differentially Private Preconditioned Stochastic Gradient Langevin Dynamics with RMSprop]
 Assume initial $\btheta_1$ is chosen independent of the data, also assume $p(\bx|\btheta)$ is $L$-smooth in $||\cdot||_2$ for any $\bx\in\mathcal{X}$ and $\btheta\in \Theta$. 
 In addition, let $\epsilon, \delta, \tau, T$ be chosen such that  $T\geq \frac{\epsilon^2 N}{32 \tau \log (2/\delta)}$. Then Algorithm  \ref{alg:psgldrms1} preserves $(\epsilon, \delta)$-differential privacy. 
\end{Thm}
\vspace{-3mm}
\begin{proof}
In every iteration, the only data access is $\bar{g}(\btheta_t;\D_t)$ and by the $L$-Lipschitz condition, the sensitivity of $\bar{g}(\btheta_t;\D_t)$ is at most $2L$ and $G(\btheta_t)$ is at least $(\frac{1}{\lambda+L})\textbf{I}$. Given
$G(\btheta_t)=\diag(\textbf{1}\oslash (\lambda\textbf{1}+\sqrt{V(\btheta_t)})$, we have
\vspace{-3mm}
{{
\begin{align*} 
V(\btheta_t) & =\sum_{i=1}^t (\alpha^{t-i}\cdot (1-\alpha)) \cdot  \left(\bar{g}(\btheta_t;\D_t)\odot\bar{g}(\btheta_t;\D_t)\right) \\
& \leq(1-\alpha) (L \boldsymbol{1} \odot L\boldsymbol{1} )\sum_{i=1}^t \alpha^{i-1}  \leq (1-\alpha) (L \boldsymbol{1} \odot L\boldsymbol{1} ) \sum_{i=1}^\infty \alpha^{i-1} \\
& \leq  (1-\alpha)  (L \boldsymbol{1} \odot L\boldsymbol{1} ) \left(\frac{1}{1-\alpha}\right) =  (L \boldsymbol{1} \odot L\boldsymbol{1}).
\end{align*}
}}
Since $(\frac{1}{\lambda+\sqrt{L\cdot L}})\textbf{I} = (\frac{1}{\lambda +L})\textbf{I} \leq G(\btheta_t)\leq \frac{1}{\lambda}\textbf{I}$, we have $\lambda\textbf{I} \leq\diag(\textbf{1}\oslash G(\btheta_t))\leq (\lambda+L)\textbf{I}$.
Note that $\diag(\textbf{1}\oslash G(\btheta_t))$ is just an inverse of $G(\btheta_t)$.
The rest part remains the same as in Wang \etal\cite{W+15}, except for removing the $\frac{N^2\eta_t^2 G^2(\btheta_t)}{4\tau^2}$ factor from the variance $\sigma^2\textbf{I}$ in the algorithm and Gaussian mechanism in order to get the essential noise that is added to $\bar{g}(\btheta_t;\D_t)$.
\end{proof}

\vspace{-2mm}
The proof includes the Lipshitz condition denoted by $L$, which is a general assumption of differential privacy on loss functions to control the impact of single data point change to the output of the algorithm \cite{S+13, B+14, W+15}.   

\smallskip
\begin{Remark}\label{rmk:dp}
The algorithm can privately release the entire sequence of parameter updates since it ensures differential privacy at each iteration. 
\end{Remark} 

Our differentially private pSGLD with RMSprop is described in Algorithm \ref{alg:psgldrms1} and we will extend it to be secure in a distributed setting via HE. 

\begin{algorithm}[thb!]
\caption{Differentially Private Preconditioned Stochastic Gradient Langevin Dynamics with RMSprop (DP-pSGLD with RMSprop)}\label{alg:psgldrms1}
\begin{algorithmic}
\State {\bfseries Input:} Data $\D,\tau,T,\gamma\in(0,1)$, RMSprop parameters $\lambda,\alpha$, privacy parameters $\epsilon, \delta,$ Lipschitz constant $L$ 
\State {\bfseries Output:} $\{\btheta_t\}_{t=1:T}$
\smallskip
\State Initialize $\bV_0\leftarrow\textbf{0}$, random $\btheta_1$ 
\For{$t=1$ {\bfseries to} $\lfloor {NT}/\tau \rfloor$}
	\State Sample a mini-batch of size $\tau, \D_t= \{\bx_{t_1},...,\bx_{t_\tau} \}$ 
	\State Estimate   $\bar{g}(\btheta_t;\D_t)=\frac{1}{\tau}\sum_{i=1}^{\tau}\nabla\log{p(\bx_{t_i}|\btheta_t})$
	\State $V(\btheta_t) \leftarrow \alpha V(\btheta_{t-1})+(1-\alpha)\bar{g}(\btheta_t;\D_t)\odot\bar{g}(\btheta_t;\D_t)$
	\State $G(\btheta_t) \leftarrow \diag(\textbf{1}\oslash (\lambda\textbf{1}+\sqrt{V(\btheta_t)})$
	\State $\eta_t \leftarrow \frac{\gamma\epsilon^2(\lambda+L)\tau}{32L^2 N\log{(\frac{2.5NT}{\tau\delta})\log{(\frac{2}{\delta } )}t}} $
	\State $\btheta_{t+1}\leftarrow \btheta_{t}
    -\frac{\eta_t}{2} \bigl[G(\btheta_{t})\bigl(\nabla_{\btheta}\log p(\btheta_{t})+ N\bar{g} (\btheta_{t};\D_t)\bigr)+\Gamma(\btheta_{t})\bigr]+\N(0,\eta_t G(\btheta_{t}))$
\EndFor
\end{algorithmic}
\end{algorithm}

\begin{Remark}\label{rmk:decompose}
The only intermediary statistics required for running this algorithm, which is a gradient, can be linearly decomposed and locally calculated at individual sites: $\bar{g}(\btheta_t;\D_t)=\frac{1}{K}\sum_{k=1}^{K} \bar{g}_k(\btheta_t;\D_t^k)$, where $K$ is the number of local sites, $\D^k=\{\bx_i^k\}$ represents local data of the $k$-th site, $\D_t^k=\{\bx^k_{t_i}\}$ is a subset of {$(\tau/K)$} data items randomly chosen from $\D^k$ at iteration $t$, and  {$\bar{g}_k(\btheta_t;\D_t^k)=\frac{K}{\tau} \sum_{i=1}^{\tau/K}\nabla\log{p(\bx^k_{t_i}|\btheta_t}$) {when $\tau$ is assumed to be divided equally across $K$ sites.}}
\end{Remark}


\subsection{Secure and Differentially Private Distributed Bayesian Learning}

In this section, we aim to explore the pSGLD algorithm with RMSprop on distributed data in a secure and privacy-preserving manner. 
We present a method to update parameters from homomorphically encrypted intermediate information while decrypting the model estimators at each iteration.
In our protocol, there are three parties: a cryptographic service provider (CSP), collaborative sites (data providers), and a cloud server. The CSP is a legitimate owner of cryptographic keys that are used for encryption ($\pk$), decryption ($\sk$), and homomorphic computation ($\evk$). 
At the beginning of the protocol, the CSP generates the key trio $(\sk, \pk, \evk)$. 
The collaborative sites use the public key to encrypt their data before outsourcing and then send encrypted data to the cloud services. 
The cloud server has only access to the evaluation key for homomorphic computation and performs a certain analysis without decrypting data.
Our protocol consists of two phases of computation: (1) a one-time preparation phase of standardizing the data matrix; (2) an iterative estimation phase of the model parameters. 

\subsubsection{Precomputation Phase}
In this phase, the data matrix is first normalized by subtracting the mean from each column and dividing each column by the standard deviation. 
Then it is divided by a quantity proportional to the maximum between its $\ell_2$ norm. 
Each local site encrypts their local sample size $n_k$ and local mean $\bar{\bx}_k= \frac{1}{n_k} \sum \bx_i^k$.
Then the server securely aggregates the encrypted local information across institutions. 
The resulting ciphertexts are decrypted with the secret key of the CSP. 
It follows from the homomorphic properties of the CKKS scheme that each ciphertext represents an approximate value to the desired result. 
Later, the CSP disseminates the global sample size $N$ and the sample sum $\bs$. 
Using the global information, the sample mean $\bar{\bx}$ can be easily computed as $\bs/N$. 
Similarly, each site can get the desired standard deviation and global maximum by the aid of the CSP. 
We adopt a random masking technique to obscure the inputs while computing their maximum, suggested in~\cite{KLOJ19}.
We refer to Appendix A for an explicit description.

\subsubsection{Iterative Model Estimation Phase}

At each iteration, the server securely aggregates the local gradients over encryption and updates the parameters while adding the Gaussian noise inside the secure computation. 
As noted in Remark~\ref{rmk:dp}, our pSGLD algorithm guarantees differential privacy of intermediate model estimators, so it allows us to decrypt and use them as fresh inputs for the subsequent iteration. 


\paragraph{Gradient estimation.} As mentioned in Remark~\ref{rmk:decompose}, the gradient can be locally calculated at individual sites. 
At each iteration, each site sends an encryption of the local gradient to the server.
For notational simplicity, we let $\bar{g}_{t}^k=\bar{g}_k(\btheta_t;\D_t^k)$ be the local gradient of the $k$-th site at iteration $t$ and denote its encrypted gradient by $\Enc(\bar{g}_{t}^k)$. 
The server aggregates the ciphertexts over encryption, yielding a ciphertext $\ct_{g_t}$ which encrypts a plaintext approximating to the global gradient $\bar{g}_{t}=\bar{g}(\btheta_t;\D_t)$. 
Here, we delay a division operation by $K$ for the sake of optimization, so we have $\Dec(\ct_{g_t})\approx K \cdot (\Delta \bar{g}_{t})$. 

\paragraph{Gradient variance.} The gradients are updated recursively by adding a fraction $\alpha$ of the update vector of the past time to the current update. This implies that at each iteration we should perform the rescaling operation right after a scalar multiplication. We note that the rescaling operation in the CKKS scheme divides an input ciphertext by a factor of $\Delta$ and a ciphertext modulus finally becomes too small to carry out further computation.  
To address this problem, we express the recursive update of $V(\btheta_t)$ as a weighted sum of the previous data:
\vspace{-2mm}
\begin{align}\label{eqn:vtheta}
{V(\btheta_t)} & \small{= \sum_{i=1}^t (\alpha^{t-i}\cdot (1-\alpha)) \cdot  \left(\bar{g}_i \odot\bar{g}_i \right).}    
\end{align}
Since the gradients are given as freshly encrypted ciphertexts (\ie they are with respect to the largest modulus), it is enough to cope with a few levels of computation, thereby achieving a better performance. 
{To be precise, suppose that at time $t$ we have encryptions of $(\bar{g}_{i} \odot \bar{g}_{i})$ for $1\le i < t$.
A naive solution for computing $V(\btheta_t)$ is to encrypt each $d$-dimensional vector $\bar{g}_{t}^k$ as a single ciphertext in a way that each entry is aligned with the vectors at the previous steps, and to evaluate the Eq.~\eqref{eqn:vtheta} by using the pure SIMD operations (addition and multiplication) on encrypted vectors without any interaction between the slots. 
In this case, we need a single multiplication and $t$ scalar multiplication for the evaluation.
We observe that a ciphertext can hold $(n/2)$ different plaintext slots for a large ring dimension $n$, and hence we can compute $r=(n/2d)$ scalar operations in parallel. 
This implies that the local gradient information of $r$ many different iterations can be packed into a single ciphertext without overlapping. 
That is, for some $t \le r$, each term of Eq.~\eqref{eqn:vtheta} is sequentially encrypted as a single ciphertext, and we finally add up all the resulting vectors by adding the output ciphertext to its rotations recursively.  
As a result, this method can be computed using a single multiplication, scalar multiplication, and $\log t$ rotations. 
In general, it only takes a single multiplication, $\ceil{t/r}$ scalar multiplications and $min\{\log r,\log t\}$ rotations.
Let $\ct_V$ denote the output ciphertext. 
}

\smallskip
\begin{Remark}
We remark that for any $\alpha \in [0,1)$, the value $\alpha^p$ goes to zero as $p$ increases. 
We may assume that the value of $\alpha^p \cdot (1-\alpha)$ is negligible for a sufficiently large $p$. 
Despite $\alpha \sim 1$, we observe that $\alpha^p \cdot (1-\alpha) < 10^{-11}$ when $\alpha=0.9$ and $p>218$. 
In practice, this gives us an approximation of the values $V(\btheta_t)$ by computing $219$ terms before from the current point $t$: 
$${V(\btheta_t) \approx \sum_{i=t-218}^{t} (\alpha^{t-i}\cdot (1-\alpha)) \cdot  \left(\bar{g}_i \odot\bar{g}_i \right).}$$
\end{Remark}


\paragraph{RMSprop preconditioner.} 
Existing HE schemes only allow the evaluation of polynomial functions, so the Taylor approximation is commonly used for approximation of analytic function. 
However, it is a local approximation near a certain point, we should use a high degree Taylor polynomial to guarantee accuracy. 
There is a global approximation method that minimizes the mean squared error proposed by Kim \etal\cite{KSW+18}, but the Chebyshev approximation is more accurate and numerically stable than this. Therefore we adopt the Chebyshev approximation of $1/(\lambda+\sqrt{x})$ for $\lambda \in \bbR$. 

Recall that the Chebyshev polynomials are defined by the recurrence relation $T_{j+1}(x)=2x \cdot T_{j}(x)-T_{j-1}(x)$ with $T_0(x)=1$ and $T_1(x)=x$. 
Note that these polynomials form an orthogonal basis over the interval $[-1,1]$. 
It can be generalized to $[a,b]$ by defining $\widetilde{T_j}(x) = T_j((2x-(b+a))/(b-a))$ for $x \in [a,b]$. 
Then $1/(\lambda+\sqrt{x})$ can be approximated by a truncated Chebyshev series $({c_0}/{2}) + \sum_{j}  c_j \cdot \widetilde{T}_j(x)$.
In order to evaluate the polynomial efficiently, we express the approximation polynomial in standard form of $\sum_j a_j x^j$. 
We observe that the plaintext of the input ciphertext $\ct_{V}$ is scaled by a factor of $K^2$, so we re-write it into the form $\sum_j (a_j/K^{2j})\cdot  (K^2\cdot x)^j$. 
Consequently, we can get an encryption of $G(\btheta_t)$ by evaluating $\sum_j (a_j/K^{2j})\cdot (\ct_{V})^j$, say the resulting ciphertext $\ct_G$.
 
\smallskip
\begin{Remark}
In practice, it is a good option to adjust the gradient $\bar{g}_t^k$ so that it is close to Lipschitz $L$ as $\tilde{g}_t^k=c\cdot \bar{g}_t^k$ with an auxiliary constant $c$ and limited to $10^{-3}$ (i.e., $\tilde{g}_t^k=\max{(\tilde{g}_t^k, 10^{-3})}$) for learning and stability of the approximation. As long as the adjusted gradient $\tilde{g}_t$ does not exceed Lipschitz $L$, Algorithm \ref{alg:psgldrms1} still satisfies $(\epsilon, \delta)$-differential privacy.
\end{Remark}


\paragraph{Parameter update.}
{In the end, the server updates the model parameters by computing the changes in the parameters as follows:}
$${\Delta\btheta_t =-\frac{\eta_{t}}{2} G(\btheta_{t}) ( \nabla_{\btheta} \log p(\btheta_t) + N \bar{g}(\btheta_t;\D_t))+ \eta'_{t} G(\btheta_t) \bz},$$

where $\bz \sim \mathcal{N}(0,\textbf{I})$ and $\eta'_{t}= \sqrt{(\lambda+L)\eta_{t}}$.  
For simplicity of the algorithm, we assume that local sites can share their prior beliefs on the parameters with no loss of privacy and utilize the same local mini-batch size. 
As noted above, since we did not divide the encryption of the global gradient by $K$, the server multiplies the ciphertext $\ct_{g_t}$ by $\frac{\eta_t N}{2K}$.
The resulting ciphertext is added by a constant  $(\frac{\eta_t }{2} \nabla_{\btheta} \log p(\btheta_t) - \eta'_t \bz)$ {since $\nabla_{\btheta} \log p(\btheta_t)$ is a function of disclosable $\btheta_t$,} and then multiplied by the ciphertext $\ct_G$.
The server finally updates the model parameters $\btheta_{t}$ using the encrypted change and sends the encrypted updated model estimator $\ct_{\btheta_{t+1}}$ to the CSP. 
After it is decrypted with the secret key of the CSP, it is sent back to the server while ensuring its privacy via DP.


\begin{algorithm}[tb]
\caption{Iterative Estimation}\label{alg:iter}
\begin{algorithmic}
\State {\bfseries Input:}  
$\tau,T,\gamma\in(0,1),\lambda,\alpha,\epsilon,\delta,L,\{a_j\}$

\State {\bfseries Output:} $\{\btheta_t\}_{t=1:T}$
\smallskip
\State Initialize random $\btheta_1$
\For{$t=1$ {\bfseries to} $\niter$}
	\State \textbf{[At local sites]:}
	\For{$k=1$ {\bfseries to} $K$}
	\State {Compute local gradient, encrypt, and transmit $\Enc(\bar{g}_t^{k})$ to the server}
	\EndFor
	\vspace{1.2mm}
	\State \textbf{[At the cloud server]:}
	\State $\ct_{g_t} \leftarrow \sum_{k=1}^{K} \Enc(\bar{g}_t^{k})$
	\State {$\ct_{V} \leftarrow \sum_{i=t-218}^{t} (\alpha^{t-i} \cdot (1-\alpha)) \cdot (\ct_{g_i} \cdot \ct_{g_i})$}
	\State $\ct_{G} \leftarrow \sum_j \frac{a_j}{K^{2j}}\cdot (\ct_{V})^j$
	\State $\eta_t \leftarrow \frac{\gamma\epsilon^2(\lambda+L)\tau}{32L^2 N\log{(\frac{2.5NT}{\tau\delta})\log{(\frac{2}{\delta } )}t}}$
	\State $\eta'_{t}\leftarrow \sqrt{(\lambda+L)\eta_t}$
	\State $\ct_{\btheta_{t+1}} \leftarrow \btheta_{t} - \bigl[\frac{\eta_{t}}{2}(\nabla_{\btheta} \log p(\btheta_t) + \frac{N}{K} \cdot \ct_{g_{t}}) - \eta'_{t}\N(0,\bI)\bigr] \cdot \ct_{G}$
	\vspace{1.2mm} 
	\State \textbf{[At the CSP]:}
	\State $\btheta_{t+1} \leftarrow \Dec(\ct_{\btheta_{t+1}})$
	\State Send $\btheta_{t+1}$ back to the server 
	\vspace{1.2mm} 
	\State \textbf{[At the cloud server]:}
	\State {Disseminate $\btheta_{t+1}$ to each local site} 
\EndFor
\end{algorithmic}
\end{algorithm}
 

\subsubsection{Threat Model}
Firstly, we assume that the cloud server is semi-honest (\ie honest but curious). 
If we ensure the semantic security of the
underlying HE scheme, all the computations on the server are processed in encrypted form, so the server learns nothing from
the encrypted data. 
Secondly, we assume that the CSP is not allowed to collude with the server or each site. 
The CSP should not be given access to data that are not part of the query from the server. 
Lastly, we assume that the local sites should not collude. 
That is, their local information (\eg local sample size, mean, or variance) is not disclosed to the other local parties even though the global information is given to all the local sites.

\section{Experiments}

In this section, we explain how to select the HE parameters and demonstrate the applicability of the proposed method on a few models.

\subsection{HE Parameter Selection}\label{sec:HEpar}
We employ the Residue Number System (RNS) variant of the CKKS scheme~\cite{CHK+18b, KSLM19}.  
Our source code is developed in C++ with Microsoft SEAL version 3.4~\cite{SEAL34}.
A freshly encrypted ciphertext of the CKKS scheme is represented as a pair of polynomials in the ring $\bbZ_Q[x]/(x^n+1)$ where $Q$ is set as a product of $(\ell+1)$ pairwise co-primes $q_i$ (\ie $Q=\prod_{i=0}^\ell q_i$). The primes $q_i$'s are chosen to have roughly the same size as the scaling factor $\Delta$. 
As noted before, we perform the rescaling operation after each multiplication, so a ciphertext at the level $i$ is scaled down by a factor of $q_i$ and the underlying plaintext is also approximately reduced by the same factor.
As a result, it turns out that the largest ciphertext modulus size is determined by the multiplicative depth of a circuit to be evaluated. 
In our protocol, it requires $\ell=3 + \ceil{(\log(d'+1)}$ where $d'$ denotes the degree of an approximation polynomial of $1/(\lambda+\sqrt{x})$. 
For $\lambda={10^{-4}}$, we used the Chebyshev approximation of degree $7$ over the interval $[10^{-3},1]$, and we therefore deduce $\ell=6$.
The scaling factor is set to $\Delta=2^{45}$~($\approx \log q_i$) and the bit-size of the largest modulus is set to $\log PQ = 380$ where $P$ is a special modulus to reduce the noise growth during homomorphic operations. 
We take the ring dimension $n= 2^{14}$ to ensure 128 bits of security against the known attacks on the LWE problem~\cite{HESecurityStandard}. 
{Details are given in Appendix B.}

\subsection{Experimental Results}
We demonstrate the feasibility of the proposed approach via several experiments on Bayesian density estimation with simple simulation, Bayesian logistic regression, and Bayesian survival analysis. In each experiment, we compare the results of models in three different settings: (1) global model without DP and HE on centralized data, (2) global model with DP on centralized data, and (3) federated model with DP and HE on distributed data. 
We note that the purpose of this experiment is to compare methods on the same model architecture but different settings. The performance of the first model is a reference value. 
We randomly selected 80\% of data for training and 20\% of data for testing and repeated ten times; their averaged results are presented. The prior is set to $p(\btheta)=\mathcal{N}(0,\sigma^2\textbf{I})$, and ${\lambda = 10^{-4}}, \alpha = 0.9, \gamma=0.99, L=1$ for simplicity. 
The privacy budget $\delta$ is is set $10^{-5}$. The step sizes for the global model with no DP and HE are determined by $\eta_t = {1.65 \times 10^{-2} /t}$ in favor of it. 
As mentioned earlier, the batch size of $\tau$ at each iteration is defined by the value in the global setting (\ie a local batch size is $\tau/K$ for the $K$-site distributed scenario). 
More intensive experimental results for various priors and local sample sizes are provided in Appendix C. 
Our experiments were conducted on a machine with Intel Xeon at 3.0 GHz with a single-thread environment, compiled with GNU C++ 7.4.0 using the `-O2' optimization setting. 

\subsubsection{Bayesian density estimation with simple simulation}
\begin{figure}[t]
\begin{center}
\hspace*{\fill}
\subfloat[{\small{Learning curves for $\theta_1$}}]{\label{fig:sd}
\includegraphics[width=0.515\columnwidth]{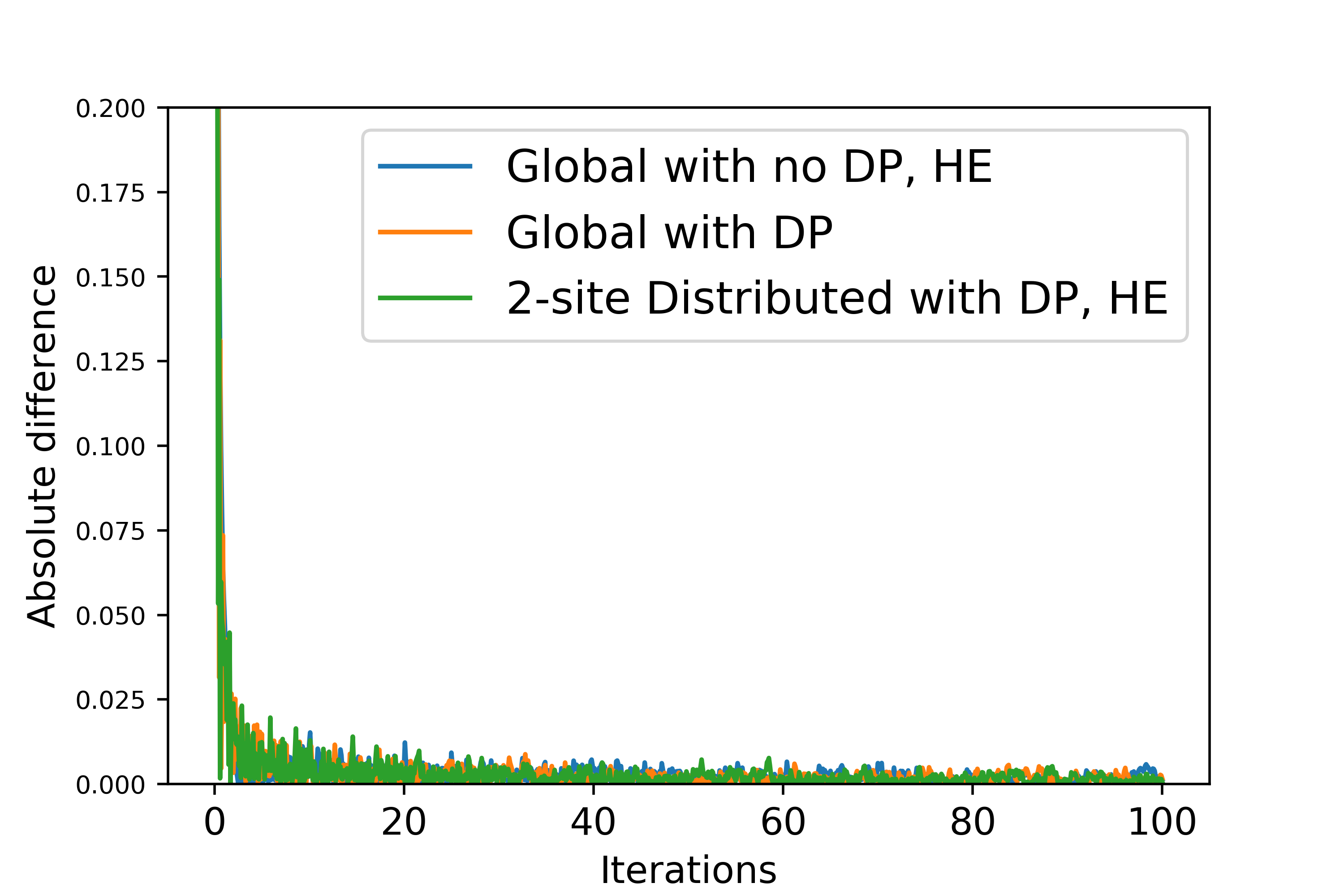}}
\subfloat[{\small{Learning curves for $\theta_2$}}]{\label{fig:sd_g}
\includegraphics[width=0.515\columnwidth]{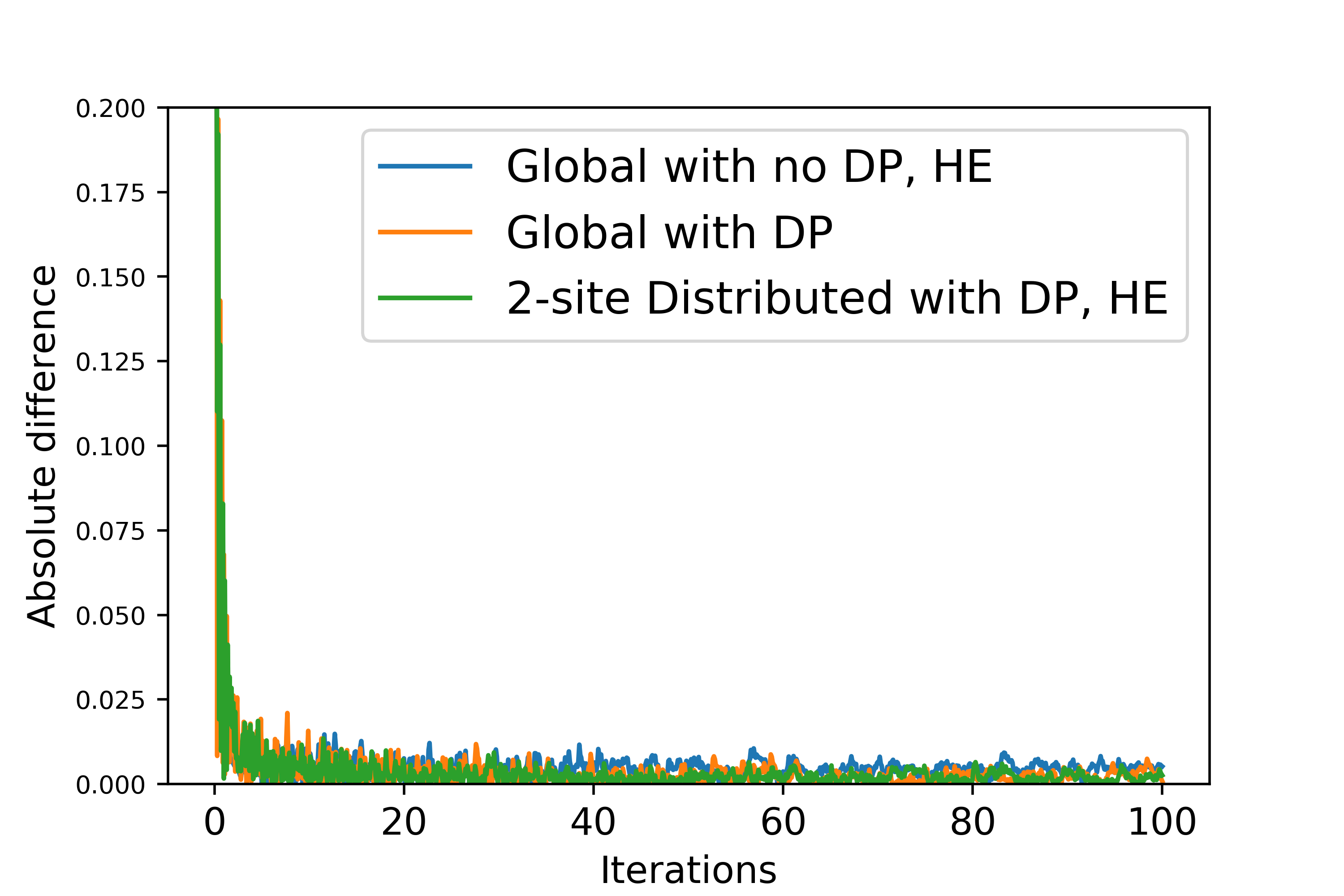}}\hfill
\hspace*{\fill}
\caption{Bayesian density estimation when $\epsilon=1$ on the artificial dataset: 2 features, 10,000 data points.}\label{fig:density}
\end{center}
\end{figure}

We implemented our approach on two-dimensional simple example data involving two parameters for the mean: 
$X_1 \sim  \N(\theta_1, 0.1); X_2 \sim  \N(\theta_2, 0.1)$ where $\theta_1 \sim \N(0,\sigma^2); \theta_2 \sim \N(0,\sigma^2)$. 
10,000 data points were drawn from the model with $\theta_1 = 0.5, \theta_2 = -0.5$, and $\sigma^2 = 1$. 
The number of sites for distributed learning is set to 2 with an equal sample size. 
Under a minibatch size of 1000 and $T=100$, we compared the results of three different models when $\epsilon=1$. Figure~\ref{fig:density} shows that the difference between the estimate from the three different models and the true value. 
The estimates from our approach seem to fluctuate but these are accurate as much as ones from the global model with and without DP on centralized data. 
This means the proposed approach enables us to learn Bayesian learning with no significant loss.

\subsubsection{Bayesian Logistic Regression}
\begin{figure}
\begin{center}
\hspace*{\fill}
\subfloat[{\small{Learning curves when $\epsilon=1$}}]{\label{fig:logreg_lraucbox}
\includegraphics[width=0.49\columnwidth]{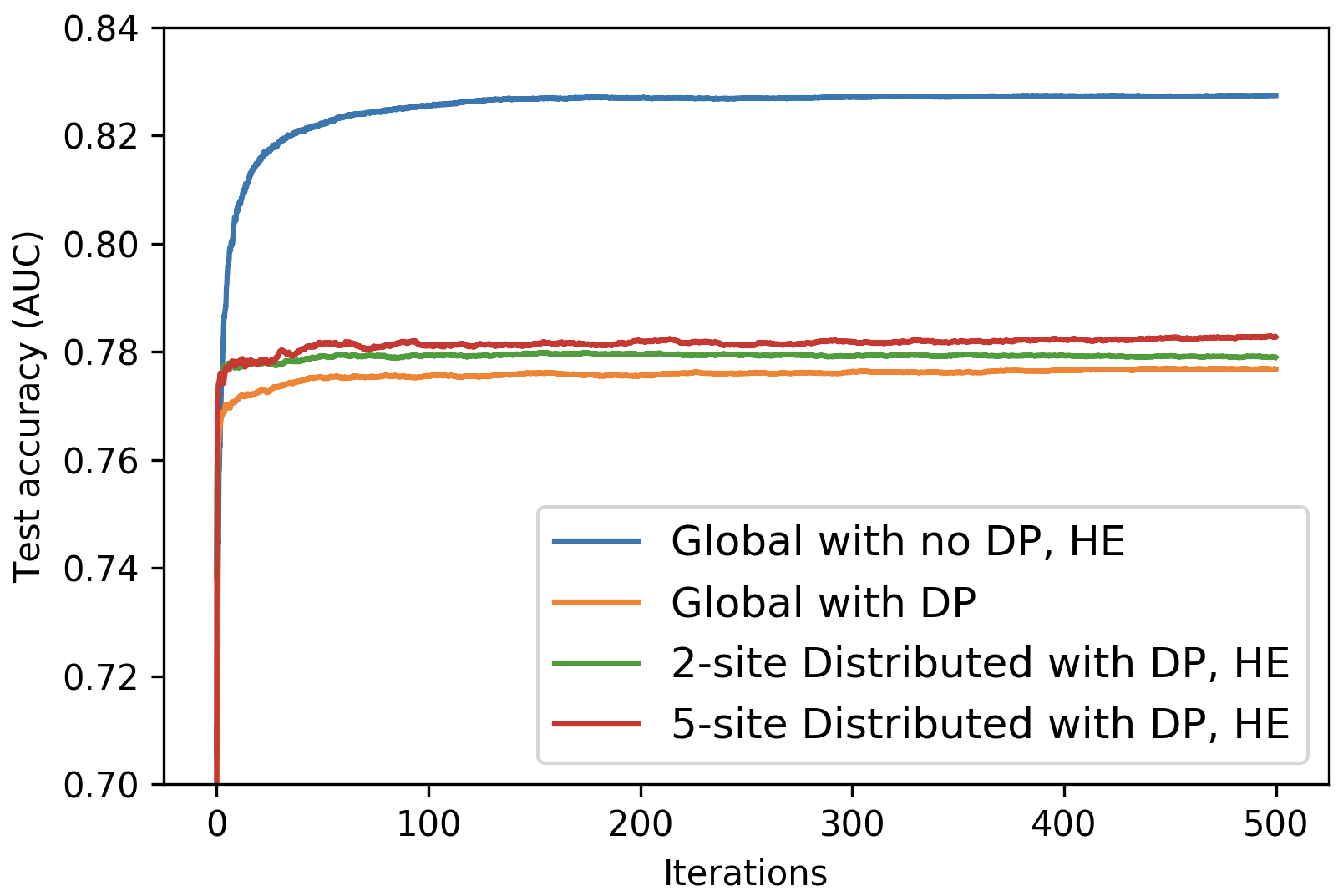}}\hfill
\subfloat[{\small{Classification accuracies}}]{\label{fig:logreg_accracy}
\includegraphics[width=0.49\columnwidth]{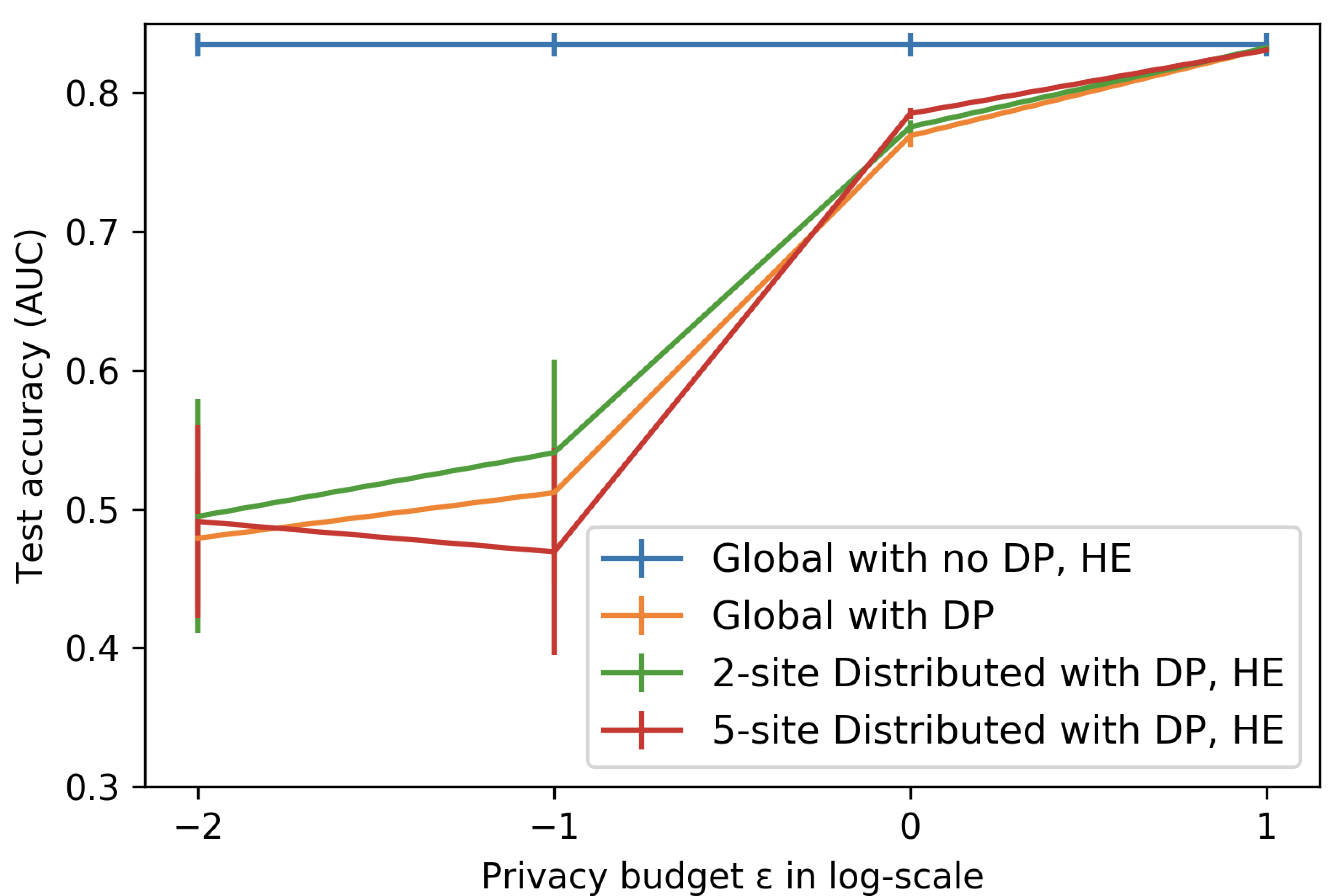}}
\hspace*{\fill}
\caption{Bayesian logistic regression on the PhysioNet dataset: 184 features, 4,000 data points.}\label{fig:logreg}
\end{center}
\end{figure}

We applied our approach to a Bayesian logistic regression model. 
We used the PhysioNet Challenge 2012 dataset~\cite{GAG+00} for binary classification to predict mortality, after pre-processing of the data. 
A Gaussian prior for each $\bbeta$ was used with the mean 0 and variance 1. Under $T=500$, a batch size of 320 at each iteration is set for both 2 and 5-site distributed scenarios. The privacy budget $\epsilon$ is changed from -2 to 1 in the log-scale. 
The results in Figure~\ref{fig:logreg} show that our approach can achieve the prediction accuracy as much as the global model with DP, which means our approach is empirically workable without any prediction accuracy loss resulting from HE and a distributed setting. 
However, due to the constraint on the learning rate of DP versions, the global model with DP including our approach cannot reach a good point with a small $\epsilon$. 
Nevertheless, the encouraging point is that prediction accuracy approaches to one of the global model without any DP and HE as $\epsilon$ increases. 

 

\subsubsection{Bayesian Survival Analysis}

\begin{figure}
\begin{center}
\hspace*{\fill}
\subfloat[{\small{Learning curves when $\epsilon=1$}}]{\label{fig:survival_lraucbox}
\includegraphics[width=0.487\columnwidth]{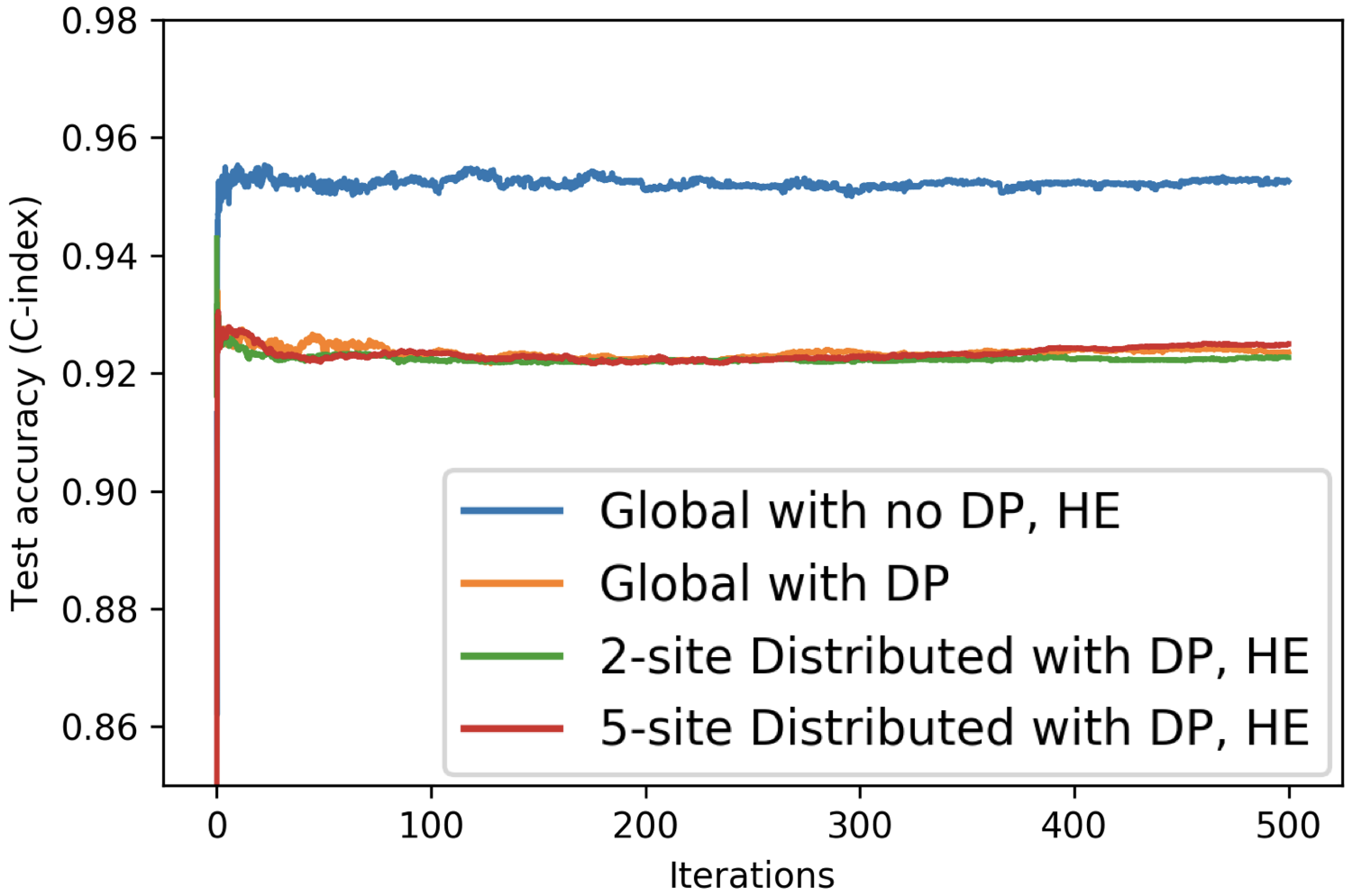}} \hfill
\subfloat[{\small{Classification accuracies}}]{\label{fig:survival_accracy}
\includegraphics[width=0.482\columnwidth]{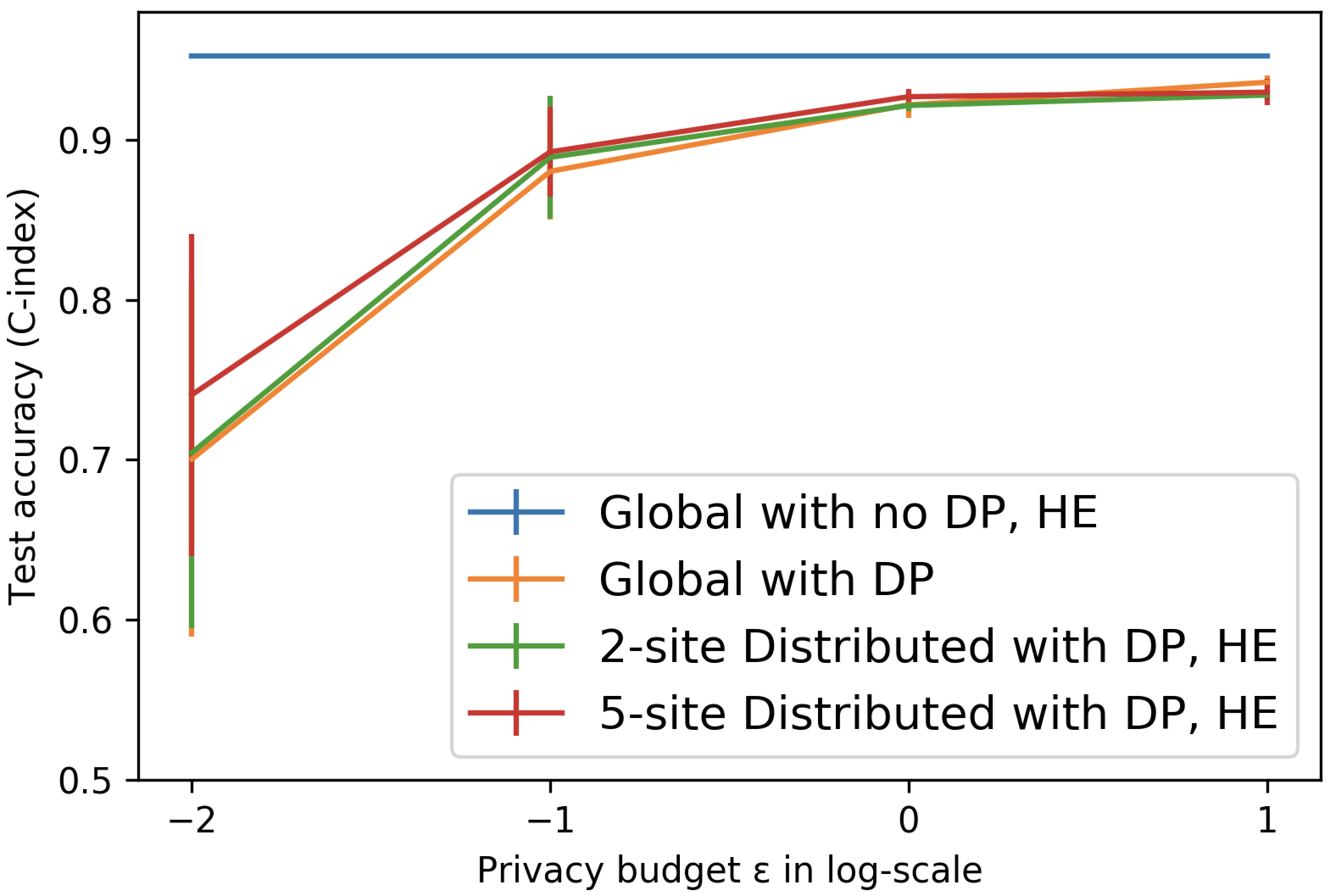}}
\hspace*{\fill}
\caption{Bayesian survival analysis on the FLchain dataset: 7 features, 7,874 data points.}
\label{fig:survival}
\end{center}
\end{figure}

Survival analysis is a class of statistical methods to study the time until the occurrence of a specified event. 
Among several models to conduct survival analysis, we focus on the Exponential distribution-based parametric survival model. The same hyper-parameter setting as logistic regression was applied here. 
We used the FLchain dataset to predict the mortality based on the association of the serum-free light chain \cite{K+06, D+09}.  
It is observed from the results shown in Figure~\ref{fig:survival} that the trends in prediction performance remain similar to those in logistic regression.

We stress that this can be easily generalized to other parametric survival models such as Weibull and Gompertz as well as Cox proportional hazard model that is a semi-parametric model~\cite{L+15}.


\subsubsection{Performance Analysis} 

In Table~\ref{tbl:timing}, we provide theoretical time complexity and experimental timing results for the aforementioned tasks. 
Here, let $\HM$ denote the homomorphic multiplication, $\SM$ the scalar multiplication, and $\Rot$ the rotation operation on the encrypted vector.
For the HE parameter setting, the key generation takes around 3.2 seconds. 
At the pre-processing phase, the complexity of all the parties is linear in the number of local sites $K$. 
We notice that it is independent of the number of features since the use of the SIMD technique allows one to perform homomorphic computation in parallel.
On the other hand, at the model parameter estimation phase, all the parties achieve a linear time complexity in the number of iterations {$T$}.

We remark that the numbers in local sites are the total running time but it might not depend on the number of participating institutions because such computation can be executed in a synchronous way. 
We see that in several models the prediction accuracy reaches a good point in a reasonable time, demonstrating the feasibility of our approach.


\begin{table*}[bt]
\caption{Time complexity analysis. Let $T'= \lfloor NT/\tau \rfloor$. Numbers are in seconds (sec) or minutes (min). We denote by $d$ and $r={8192}/{d}$ the number of features and the maximal number of $d$-dimensional vectors which are capable of being encrypted in a single ciphertext, respectively. }\label{tbl:timing}
\vspace{-2mm}
\renewcommand{\arraystretch}{1.21}  
\renewcommand{\tabcolsep}{0.8mm}   
\begin{center}
\begin{scriptsize}
\begin{tabular}{ccccccccc}
\toprule
\multirow{2}{*}{\textbf{{Task}}}& \textbf{{Number}} & \textbf{{Number}} & \multicolumn{3}{c}{\textbf{{Precomputation}}}  & \multicolumn{3}{c}{\textbf{{Iterative model estimation}}} \\ \cmidrule(lr){4-6} \cmidrule(lr){7-9} 
& \textbf{{of sites}} & \textbf{{of iters}} & \textbf{{ Local }} & \textbf{{Server}} & \textbf{{CSP}}  & \textbf{{ Local }} & \textbf{{Server}} & \textbf{{CSP}} \\
\midrule
 {Theoretical} & \multirow{3}{*}{$K$} & \multirow{3}{*}{$T$} &\multirow{3}{*}{$4K \Enc$} & ${K \HM} $ &  $1 \Enc$
& \multirow{3}{*}{$KT' \Enc$} & $\left(T' \cdot (8 + \ceil{\frac{219}{r}})\right)\SM +$ & \multirow{3}{*}{$T' \Dec$} \\
{Complexity} & & & & $+ (K+1) \SM$ & $+(K+4) \Dec$  & & $ (9T') \HM +$ & \\
 & & & & &  & & $ \left(T' \cdot min\{{\ceil{\log t},} 8, \ceil{\log r}\}\right) \Rot$ & \\

\midrule
{Density } & \multirow{2}{*}{2} &\multirow{2}{*}{100} & \multirow{2}{*}{0.05 sec} & \multirow{2}{*}{0.04 sec} &  \multirow{2}{*}{0.09 sec} & \multirow{2}{*}{34.4 sec} & \multirow{2}{*}{4.2 min} & \multirow{2}{*}{1.0 sec} \\
{estimation} & &  &  &  &  &   &  & \\
\midrule
{Logistic} & 2 & \multirow{2}{*}{500} & 0.07 sec & 0.2 sec & 0.4 sec &  7.0 min  & 34.1 min  & 9.4 sec  \\
{regression} & 5 & & 0.1 sec & 0.4 sec & 0.2 sec & 15.6 min & 36.5 min  & 10.4 sec \\
\midrule
{Survival} & 2 & \multirow{2}{*}{500} & 0.07 sec & 0.2 sec & 0.1 sec & 14.2 min & 67.1 min  & 17.3 sec  \\
{analysis} & 5 & & 0.09 sec & 0.4 sec & 0.1 sec & 28.6 min & 69.7 min & 17.7 sec  \\
\bottomrule
\end{tabular}
\end{scriptsize}
\end{center}
\vskip -0.1in
\end{table*}

\section{Conclusion}
{This paper presents the first work to combine DP and HE for a popular Bayesian optimization framework pSGLD in a distributed setting. 
We demonstrated its applicability to the generalized linear model (\eg logistic regression) and the time-to-event model (\eg survival models). 
It can be further extended to Bayesian deep learning models to address the dilemma of privacy and utility, closing the technology gap in data scarcity and model generalizability. 
We provide privacy and security analysis with provable guarantees to this optimization strategy. 
This might address a lot of challenges in healthcare, education, and finance disciplines, in which data are highly confidential and hard to acquire from distributed owners.}

\bibliography{ref}

\providecommand{\latin}[1]{#1}
\makeatletter
\providecommand{\doi}
  {\begingroup\let\do\@makeother\dospecials
  \catcode`\{=1 \catcode`\}=2 \doi@aux}
\providecommand{\doi@aux}[1]{\endgroup\texttt{#1}}
\makeatother
\providecommand*\mcitethebibliography{\thebibliography}
\csname @ifundefined\endcsname{endmcitethebibliography}
  {\let\endmcitethebibliography\endthebibliography}{}
\begin{mcitethebibliography}{41}
\providecommand*\natexlab[1]{#1}
\providecommand*\mciteSetBstSublistMode[1]{}
\providecommand*\mciteSetBstMaxWidthForm[2]{}
\providecommand*\mciteBstWouldAddEndPuncttrue
  {\def\EndOfBibitem{\unskip.}}
\providecommand*\mciteBstWouldAddEndPunctfalse
  {\let\EndOfBibitem\relax}
\providecommand*\mciteSetBstMidEndSepPunct[3]{}
\providecommand*\mciteSetBstSublistLabelBeginEnd[3]{}
\providecommand*\EndOfBibitem{}
\mciteSetBstSublistMode{f}
\mciteSetBstMaxWidthForm{subitem}{(\alph{mcitesubitemcount})}
\mciteSetBstSublistLabelBeginEnd
  {\mcitemaxwidthsubitemform\space}
  {\relax}
  {\relax}

\bibitem[Fredrikson \latin{et~al.}(2014)Fredrikson, Lantz, Jha, Lin, Page, and
  Ristenpart]{F+14}
Fredrikson,~M.; Lantz,~E.; Jha,~S.; Lin,~S.; Page,~D.; Ristenpart,~T. Privacy
  in pharmacogenetics: An end-to-end case study of personalized warfarin
  dosing. 23rd {USENIX} Security Symposium. 2014; pp 17--32\relax
\mciteBstWouldAddEndPuncttrue
\mciteSetBstMidEndSepPunct{\mcitedefaultmidpunct}
{\mcitedefaultendpunct}{\mcitedefaultseppunct}\relax
\EndOfBibitem
\bibitem[Shokri \latin{et~al.}(2017)Shokri, Stronati, Song, and
  Shmatikov]{S+17}
Shokri,~R.; Stronati,~M.; Song,~C.; Shmatikov,~V. Membership inference attacks
  against machine learning models. 2017 IEEE Symposium on Security and Privacy
  (SP). 2017; pp 3--18\relax
\mciteBstWouldAddEndPuncttrue
\mciteSetBstMidEndSepPunct{\mcitedefaultmidpunct}
{\mcitedefaultendpunct}{\mcitedefaultseppunct}\relax
\EndOfBibitem
\bibitem[Dwork(2006)]{Dwork06}
Dwork,~C. Differential Privacy. 33rd International Colloquium on Automata,
  Languages and Programming, part {II} ({ICALP} 2006). Venice, Italy, 2006; pp
  1--12\relax
\mciteBstWouldAddEndPuncttrue
\mciteSetBstMidEndSepPunct{\mcitedefaultmidpunct}
{\mcitedefaultendpunct}{\mcitedefaultseppunct}\relax
\EndOfBibitem
\bibitem[Dwork \latin{et~al.}(2006)Dwork, McSherry, Nissim, and
  Smith]{Dwork2006-sx}
Dwork,~C.; McSherry,~F.; Nissim,~K.; Smith,~A. \emph{Theory of cryptography};
  Springer, 2006; pp 265--284\relax
\mciteBstWouldAddEndPuncttrue
\mciteSetBstMidEndSepPunct{\mcitedefaultmidpunct}
{\mcitedefaultendpunct}{\mcitedefaultseppunct}\relax
\EndOfBibitem
\bibitem[Gentry(2009)]{Gen09}
Gentry,~C. A fully homomorphic encryption scheme. Ph.D.\ thesis, Stanford
  University, 2009\relax
\mciteBstWouldAddEndPuncttrue
\mciteSetBstMidEndSepPunct{\mcitedefaultmidpunct}
{\mcitedefaultendpunct}{\mcitedefaultseppunct}\relax
\EndOfBibitem
\bibitem[Aono \latin{et~al.}(2016)Aono, Hayashi, Trieu~Phong, and Wang]{A+16}
Aono,~Y.; Hayashi,~T.; Trieu~Phong,~L.; Wang,~L. Scalable and secure logistic
  regression via homomorphic encryption. Proceedings of the 6th ACM Conference
  on Data and Application Security and Privacy. 2016; pp 142--144\relax
\mciteBstWouldAddEndPuncttrue
\mciteSetBstMidEndSepPunct{\mcitedefaultmidpunct}
{\mcitedefaultendpunct}{\mcitedefaultseppunct}\relax
\EndOfBibitem
\bibitem[Kim \latin{et~al.}(2019)Kim, Lee, Ohno-Machado, and Jiang]{KLOJ19}
Kim,~M.; Lee,~J.; Ohno-Machado,~L.; Jiang,~X. Secure and Differentially Private
  Logistic Regression for Horizontally Distributed Data. \emph{IEEE
  Transactions on Information Forensics and Security} \textbf{2019}, \emph{15},
  695--710\relax
\mciteBstWouldAddEndPuncttrue
\mciteSetBstMidEndSepPunct{\mcitedefaultmidpunct}
{\mcitedefaultendpunct}{\mcitedefaultseppunct}\relax
\EndOfBibitem
\bibitem[Robbins and Monro(1951)Robbins, and Monro]{RM51}
Robbins,~H.; Monro,~S. A stochastic approximation method. \emph{The annals of
  mathematical statistics} \textbf{1951}, 400--407\relax
\mciteBstWouldAddEndPuncttrue
\mciteSetBstMidEndSepPunct{\mcitedefaultmidpunct}
{\mcitedefaultendpunct}{\mcitedefaultseppunct}\relax
\EndOfBibitem
\bibitem[Neal(1995)]{N95}
Neal,~R.~M. Bayesian learning for neural networks. Ph.D.\ thesis, University of
  Toronto, 1995\relax
\mciteBstWouldAddEndPuncttrue
\mciteSetBstMidEndSepPunct{\mcitedefaultmidpunct}
{\mcitedefaultendpunct}{\mcitedefaultseppunct}\relax
\EndOfBibitem
\bibitem[Ahn \latin{et~al.}(2012)Ahn, Korattikara, and Welling]{A+12}
Ahn,~S.; Korattikara,~A.; Welling,~M. Bayesian posterior sampling via
  stochastic gradient fisher scoring. Proceedings of the 29th International
  Coference on International Conference on Machine Learning. 2012; pp
  1771--1778\relax
\mciteBstWouldAddEndPuncttrue
\mciteSetBstMidEndSepPunct{\mcitedefaultmidpunct}
{\mcitedefaultendpunct}{\mcitedefaultseppunct}\relax
\EndOfBibitem
\bibitem[Chen \latin{et~al.}(2014)Chen, Fox, and Guestrin]{C+14}
Chen,~T.; Fox,~E.; Guestrin,~C. Stochastic {G}radient {H}amiltonian {M}onte
  {C}arlo. Proceedings of the 31st International Coference on International
  Conference on Machine Learning. 2014; pp 1683--1691\relax
\mciteBstWouldAddEndPuncttrue
\mciteSetBstMidEndSepPunct{\mcitedefaultmidpunct}
{\mcitedefaultendpunct}{\mcitedefaultseppunct}\relax
\EndOfBibitem
\bibitem[Ding \latin{et~al.}(2014)Ding, Fang, Babbush, Chen, Skeel, and
  Neven]{D+14}
Ding,~N.; Fang,~Y.; Babbush,~R.; Chen,~C.; Skeel,~R.~D.; Neven,~H. Bayesian
  sampling using stochastic gradient thermostats. Advances in neural
  information processing systems. 2014; pp 3203--3211\relax
\mciteBstWouldAddEndPuncttrue
\mciteSetBstMidEndSepPunct{\mcitedefaultmidpunct}
{\mcitedefaultendpunct}{\mcitedefaultseppunct}\relax
\EndOfBibitem
\bibitem[Welling and Teh(2011)Welling, and Teh]{WT11}
Welling,~M.; Teh,~Y.~W. Bayesian learning via stochastic gradient {L}angevin
  dynamics. Proceedings of the 28th International Conference on Machine
  Learning. 2011; pp 681--688\relax
\mciteBstWouldAddEndPuncttrue
\mciteSetBstMidEndSepPunct{\mcitedefaultmidpunct}
{\mcitedefaultendpunct}{\mcitedefaultseppunct}\relax
\EndOfBibitem
\bibitem[Li \latin{et~al.}(2016)Li, Chen, Carlson, and Carin]{Li+16}
Li,~C.; Chen,~C.; Carlson,~D.; Carin,~L. Preconditioned stochastic gradient
  {L}angevin dynamics for deep neural networks. 30th AAAI Conference on
  Artificial Intelligence. 2016; pp 1788--1794\relax
\mciteBstWouldAddEndPuncttrue
\mciteSetBstMidEndSepPunct{\mcitedefaultmidpunct}
{\mcitedefaultendpunct}{\mcitedefaultseppunct}\relax
\EndOfBibitem
\bibitem[Wang \latin{et~al.}(2015)Wang, Fienberg, and Smola]{W+15}
Wang,~Y.-X.; Fienberg,~S.; Smola,~A. Privacy for free: {P}osterior sampling and
  stochastic gradient {M}onte {C}arlo. Proceedings of the 32nd International
  Coference on International Conference on Machine Learning. 2015; pp
  2493--2502\relax
\mciteBstWouldAddEndPuncttrue
\mciteSetBstMidEndSepPunct{\mcitedefaultmidpunct}
{\mcitedefaultendpunct}{\mcitedefaultseppunct}\relax
\EndOfBibitem
\bibitem[Li \latin{et~al.}(2019)Li, Chen, Liu, and Carin]{Li+19}
Li,~B.; Chen,~C.; Liu,~H.; Carin,~L. On Connecting Stochastic Gradient {MCMC}
  and Differential Privacy. The 22nd International Conference on Artificial
  Intelligence and Statistics. 2019; pp 557--566\relax
\mciteBstWouldAddEndPuncttrue
\mciteSetBstMidEndSepPunct{\mcitedefaultmidpunct}
{\mcitedefaultendpunct}{\mcitedefaultseppunct}\relax
\EndOfBibitem
\bibitem[Heikkil{\"a} \latin{et~al.}(2017)Heikkil{\"a}, Lagerspetz, Kaski,
  Shimizu, Tarkoma, and Honkela]{H+17}
Heikkil{\"a},~M.; Lagerspetz,~E.; Kaski,~S.; Shimizu,~K.; Tarkoma,~S.;
  Honkela,~A. Differentially private Bayesian learning on distributed data.
  Advances in neural information processing systems. 2017; pp 3226--3235\relax
\mciteBstWouldAddEndPuncttrue
\mciteSetBstMidEndSepPunct{\mcitedefaultmidpunct}
{\mcitedefaultendpunct}{\mcitedefaultseppunct}\relax
\EndOfBibitem
\bibitem[Cheon \latin{et~al.}(2018)Cheon, Han, Kim, Kim, and Song]{CHK+18a}
Cheon,~J.~H.; Han,~K.; Kim,~A.; Kim,~M.; Song,~Y. Bootstrapping for approximate
  homomorphic encryption. Annual International Conference on the Theory and
  Applications of Cryptographic Techniques. 2018; pp 360--384\relax
\mciteBstWouldAddEndPuncttrue
\mciteSetBstMidEndSepPunct{\mcitedefaultmidpunct}
{\mcitedefaultendpunct}{\mcitedefaultseppunct}\relax
\EndOfBibitem
\bibitem[Dwork and Roth(2014)Dwork, and Roth]{DR14}
Dwork,~C.; Roth,~A. The algorithmic foundations of differential privacy.
  \emph{Foundations and Trends{\textregistered} in Theoretical Computer
  Science} \textbf{2014}, \emph{9}, 211--407\relax
\mciteBstWouldAddEndPuncttrue
\mciteSetBstMidEndSepPunct{\mcitedefaultmidpunct}
{\mcitedefaultendpunct}{\mcitedefaultseppunct}\relax
\EndOfBibitem
\bibitem[Dwork \latin{et~al.}(2010)Dwork, Rothblum, and Vadhan]{DRV10}
Dwork,~C.; Rothblum,~G.~N.; Vadhan,~S. Boosting and differential privacy. 2010
  IEEE 51st Annual Symposium on Foundations of Computer Science. 2010; pp
  51--60\relax
\mciteBstWouldAddEndPuncttrue
\mciteSetBstMidEndSepPunct{\mcitedefaultmidpunct}
{\mcitedefaultendpunct}{\mcitedefaultseppunct}\relax
\EndOfBibitem
\bibitem[Naehrig \latin{et~al.}(2011)Naehrig, Lauter, and
  Vaikuntanathan]{LNV11}
Naehrig,~M.; Lauter,~K.; Vaikuntanathan,~V. Can homomorphic encryption be
  practical? Proceedings of the 3rd ACM workshop on Cloud computing security
  workshop. 2011; pp 113--124\relax
\mciteBstWouldAddEndPuncttrue
\mciteSetBstMidEndSepPunct{\mcitedefaultmidpunct}
{\mcitedefaultendpunct}{\mcitedefaultseppunct}\relax
\EndOfBibitem
\bibitem[Gilad-Bachrach \latin{et~al.}(2016)Gilad-Bachrach, Dowlin, Laine,
  Lauter, Naehrig, and Wernsing]{Cryptonets}
Gilad-Bachrach,~R.; Dowlin,~N.; Laine,~K.; Lauter,~K.; Naehrig,~M.;
  Wernsing,~J. Cryptonets: Applying neural networks to encrypted data with high
  throughput and accuracy. International Conference on Machine Learning. 2016;
  pp 201--210\relax
\mciteBstWouldAddEndPuncttrue
\mciteSetBstMidEndSepPunct{\mcitedefaultmidpunct}
{\mcitedefaultendpunct}{\mcitedefaultseppunct}\relax
\EndOfBibitem
\bibitem[Jiang \latin{et~al.}(2018)Jiang, Kim, Lauter, and Song]{JKLS18}
Jiang,~X.; Kim,~M.; Lauter,~K.; Song,~Y. Secure outsourced matrix computation
  and application to neural networks. Proceedings of the 2018 ACM SIGSAC
  Conference on Computer and Communications Security. 2018; pp 1209--1222\relax
\mciteBstWouldAddEndPuncttrue
\mciteSetBstMidEndSepPunct{\mcitedefaultmidpunct}
{\mcitedefaultendpunct}{\mcitedefaultseppunct}\relax
\EndOfBibitem
\bibitem[Lou and Jiang(2019)Lou, and Jiang]{SHE}
Lou,~Q.; Jiang,~L. {SHE}: A Fast and Accurate Deep Neural Network for Encrypted
  Data. Advances in Neural Information Processing Systems. 2019; pp
  10035--10043\relax
\mciteBstWouldAddEndPuncttrue
\mciteSetBstMidEndSepPunct{\mcitedefaultmidpunct}
{\mcitedefaultendpunct}{\mcitedefaultseppunct}\relax
\EndOfBibitem
\bibitem[Brutzkus \latin{et~al.}(2019)Brutzkus, Gilad-Bachrach, and
  Elisha]{BGR19}
Brutzkus,~A.; Gilad-Bachrach,~R.; Elisha,~O. Low Latency Privacy Preserving
  Inference. Proceedings of the 36th International Coference on International
  Conference on Machine Learning. 2019; pp 812--821\relax
\mciteBstWouldAddEndPuncttrue
\mciteSetBstMidEndSepPunct{\mcitedefaultmidpunct}
{\mcitedefaultendpunct}{\mcitedefaultseppunct}\relax
\EndOfBibitem
\bibitem[Cheon \latin{et~al.}(2017)Cheon, Kim, Kim, and Song]{CKKS17}
Cheon,~J.~H.; Kim,~A.; Kim,~M.; Song,~Y. Homomorphic encryption for arithmetic
  of approximate numbers. Advances in Cryptology--ASIACRYPT 2017: 23rd
  International Conference on the Theory and Application of Cryptology and
  Information Security. 2017; pp 409--437\relax
\mciteBstWouldAddEndPuncttrue
\mciteSetBstMidEndSepPunct{\mcitedefaultmidpunct}
{\mcitedefaultendpunct}{\mcitedefaultseppunct}\relax
\EndOfBibitem
\bibitem[Kim \latin{et~al.}(2018)Kim, Song, Wang, Xia, and Jiang]{KSW+18}
Kim,~M.; Song,~Y.; Wang,~S.; Xia,~Y.; Jiang,~X. Secure Logistic Regression
  Based on Homomorphic Encryption: Design and Evaluation. \emph{JMIR medical
  informatics} \textbf{2018}, \emph{6}\relax
\mciteBstWouldAddEndPuncttrue
\mciteSetBstMidEndSepPunct{\mcitedefaultmidpunct}
{\mcitedefaultendpunct}{\mcitedefaultseppunct}\relax
\EndOfBibitem
\bibitem[Kim \latin{et~al.}(2018)Kim, Song, Kim, Lee, and Cheon]{KSK+18}
Kim,~A.; Song,~Y.; Kim,~M.; Lee,~K.; Cheon,~J.~H. Logistic Regression Model
  Training based on the Approximate Homomorphic Encryption. \emph{BMC Medical
  Genomics} \textbf{2018}, \emph{11}, 83\relax
\mciteBstWouldAddEndPuncttrue
\mciteSetBstMidEndSepPunct{\mcitedefaultmidpunct}
{\mcitedefaultendpunct}{\mcitedefaultseppunct}\relax
\EndOfBibitem
\bibitem[Duchi \latin{et~al.}(2011)Duchi, Hazan, and Singer]{J+11}
Duchi,~J.; Hazan,~E.; Singer,~Y. Adaptive subgradient methods for online
  learning and stochastic optimization. \emph{Journal of machine learning
  research} \textbf{2011}, \emph{12}, 2121--2159\relax
\mciteBstWouldAddEndPuncttrue
\mciteSetBstMidEndSepPunct{\mcitedefaultmidpunct}
{\mcitedefaultendpunct}{\mcitedefaultseppunct}\relax
\EndOfBibitem
\bibitem[Kingma and Ba(2014)Kingma, and Ba]{KB15}
Kingma,~D.~P.; Ba,~J. Adam: A method for stochastic optimization. \emph{arXiv
  preprint arXiv:1412.6980} \textbf{2014}, \relax
\mciteBstWouldAddEndPunctfalse
\mciteSetBstMidEndSepPunct{\mcitedefaultmidpunct}
{}{\mcitedefaultseppunct}\relax
\EndOfBibitem
\bibitem[Tieleman and Hinton(2012)Tieleman, and Hinton]{TH12}
Tieleman,~T.; Hinton,~G. Lecture 6.5-rmsprop: Divide the gradient by a running
  average of its recent magnitude. \emph{COURSERA: Neural networks for machine
  learning} \textbf{2012}, \emph{4}, 26--31\relax
\mciteBstWouldAddEndPuncttrue
\mciteSetBstMidEndSepPunct{\mcitedefaultmidpunct}
{\mcitedefaultendpunct}{\mcitedefaultseppunct}\relax
\EndOfBibitem
\bibitem[Song \latin{et~al.}(2013)Song, Chaudhuri, and Sarwate]{S+13}
Song,~S.; Chaudhuri,~K.; Sarwate,~A.~D. Stochastic gradient descent with
  differentially private updates. 2013 IEEE Global Conference on Signal and
  Information Processing. 2013; pp 245--248\relax
\mciteBstWouldAddEndPuncttrue
\mciteSetBstMidEndSepPunct{\mcitedefaultmidpunct}
{\mcitedefaultendpunct}{\mcitedefaultseppunct}\relax
\EndOfBibitem
\bibitem[Bassily \latin{et~al.}(2014)Bassily, Smith, and Thakurta]{B+14}
Bassily,~R.; Smith,~A.; Thakurta,~A. Private empirical risk minimization:
  Efficient algorithms and tight error bounds. 2014 IEEE 55th Annual Symposium
  on Foundations of Computer Science. 2014; pp 464--473\relax
\mciteBstWouldAddEndPuncttrue
\mciteSetBstMidEndSepPunct{\mcitedefaultmidpunct}
{\mcitedefaultendpunct}{\mcitedefaultseppunct}\relax
\EndOfBibitem
\bibitem[Cheon \latin{et~al.}(2018)Cheon, Han, Kim, Kim, and Song]{CHK+18b}
Cheon,~J.~H.; Han,~K.; Kim,~A.; Kim,~M.; Song,~Y. A Full {RNS} Variant of
  Approximate Homomorphic Encryption. Selected Areas in Cryptography -- SAC
  2018. 2018; pp 347--368\relax
\mciteBstWouldAddEndPuncttrue
\mciteSetBstMidEndSepPunct{\mcitedefaultmidpunct}
{\mcitedefaultendpunct}{\mcitedefaultseppunct}\relax
\EndOfBibitem
\bibitem[Kim \latin{et~al.}(2019)Kim, Song, Li, and Micciancio]{KSLM19}
Kim,~M.; Song,~Y.; Li,~B.; Micciancio,~D. Semi-parallel logistic regression for
  {GWAS} on encrypted data. Cryptology ePrint Archive, Report 2019/294, 2019;
  \url{https://eprint.iacr.org/2019/294}\relax
\mciteBstWouldAddEndPuncttrue
\mciteSetBstMidEndSepPunct{\mcitedefaultmidpunct}
{\mcitedefaultendpunct}{\mcitedefaultseppunct}\relax
\EndOfBibitem
\bibitem[SEAL(2019)]{SEAL34}
{M}icrosoft {SEAL} (release 3.4). \url{https://github.com/Microsoft/SEAL},
  2019; Microsoft Research, Redmond, WA.\relax
\mciteBstWouldAddEndPunctfalse
\mciteSetBstMidEndSepPunct{\mcitedefaultmidpunct}
{}{\mcitedefaultseppunct}\relax
\EndOfBibitem
\bibitem[Albrecht \latin{et~al.}(2018)Albrecht, Chase, Chen, Ding, Goldwasser,
  Gorbunov, Halevi, Hoffstein, Laine, Lauter, Lokam, Micciancio, Moody,
  Morrison, Sahai, and Vaikuntanathan]{HESecurityStandard}
Albrecht,~M. \latin{et~al.}  \emph{Homomorphic Encryption Security Standard};
  2018\relax
\mciteBstWouldAddEndPuncttrue
\mciteSetBstMidEndSepPunct{\mcitedefaultmidpunct}
{\mcitedefaultendpunct}{\mcitedefaultseppunct}\relax
\EndOfBibitem
\bibitem[Golberger \latin{et~al.}(2000)Golberger, Amaral, Glass, Hausdorff,
  Ivanov, Mark, Mietus, Moody, Chung-Kan, and Stenley]{GAG+00}
Golberger,~A.; Amaral,~L.; Glass,~L.; Hausdorff,~J.~M.; Ivanov,~P.~C.;
  Mark,~R.; Mietus,~J.; Moody,~G.; Chung-Kan,~P.; Stenley,~H. PhysioBank,
  PhysioToolkit, and PhysioNet: Component of a New Research Resource for
  Complex Physiologic Signals. \emph{Circulation} \textbf{2000}, \emph{101},
  e215--e220\relax
\mciteBstWouldAddEndPuncttrue
\mciteSetBstMidEndSepPunct{\mcitedefaultmidpunct}
{\mcitedefaultendpunct}{\mcitedefaultseppunct}\relax
\EndOfBibitem
\bibitem[Kyle \latin{et~al.}(2006)Kyle, Therneau, Rajkumar, Larson, Plevak,
  Offord, Dispenzieri, Katzmann, and Melton~III]{K+06}
Kyle,~R.~A.; Therneau,~T.~M.; Rajkumar,~S.~V.; Larson,~D.~R.; Plevak,~M.~F.;
  Offord,~J.~R.; Dispenzieri,~A.; Katzmann,~J.~A.; Melton~III,~L.~J. Prevalence
  of monoclonal gammopathy of undetermined significance. \emph{New England
  Journal of Medicine} \textbf{2006}, \emph{354}, 1362--1369\relax
\mciteBstWouldAddEndPuncttrue
\mciteSetBstMidEndSepPunct{\mcitedefaultmidpunct}
{\mcitedefaultendpunct}{\mcitedefaultseppunct}\relax
\EndOfBibitem
\bibitem[Dispenzieri \latin{et~al.}(2009)Dispenzieri, Kyle, Merlini, Miguel,
  Ludwig, Hajek, Palumbo, Jagannath, Blad{\'e}, Lonial, \latin{et~al.}
  others]{D+09}
Dispenzieri,~A.; Kyle,~R.; Merlini,~G.; Miguel,~J.; Ludwig,~H.; Hajek,~R.;
  Palumbo,~A.; Jagannath,~S.; Blad{\'e},~J.; Lonial,~S., \latin{et~al.}
  International Myeloma Working Group guidelines for serum-free light chain
  analysis in multiple myeloma and related disorders. \emph{Leukemia}
  \textbf{2009}, \emph{23}, 215--224\relax
\mciteBstWouldAddEndPuncttrue
\mciteSetBstMidEndSepPunct{\mcitedefaultmidpunct}
{\mcitedefaultendpunct}{\mcitedefaultseppunct}\relax
\EndOfBibitem
\bibitem[Lu \latin{et~al.}(2015)Lu, Wang, Ji, Wu, Xiong, Jiang, and
  Ohno-Machado]{L+15}
Lu,~C.-L.; Wang,~S.; Ji,~Z.; Wu,~Y.; Xiong,~L.; Jiang,~X.; Ohno-Machado,~L.
  WebDISCO: a web service for distributed cox model learning without
  patient-level data sharing. \emph{Journal of the American Medical Informatics
  Association} \textbf{2015}, \emph{22}, 1212--1219\relax
\mciteBstWouldAddEndPuncttrue
\mciteSetBstMidEndSepPunct{\mcitedefaultmidpunct}
{\mcitedefaultendpunct}{\mcitedefaultseppunct}\relax
\EndOfBibitem
\end{mcitethebibliography}
\bibliographystyle{achemso}


\appendix


\section{Algorithm Details of Precomputation}~\label{append:pre}

\vspace{-5mm}
\subsection{{Standardization}}\label{append:pre1}

Algorithm~\ref{alg:pre1} provides the detail about our secure protocol for data standardization. In this protocol, the data matrix is securely normalized by subtracting the mean from each column and dividing each column by the standard deviation.

Suppose that $\D^k=\{\bx_i^k\}_{1\le i \le n_k}$ represents local data of the $k$-th site. 
Each local site encrypts their local sample size $n_k$ and local mean $\bar{\bx}_k= \frac{1}{n_k} \sum \bx_i^k$, and sends the encryptions to the server. 
Next, the server securely aggregates the encrypted local information across institutions. 
After that, the resulting ciphertexts are decrypted with the secret key of the CSP. 
In particular, the global sample size $N$ is an integer, so if we take a sufficiently large scale factor of $\Delta$, the decryption result is very close to $N$ and we can get the exact $N$ by computing the nearest integer of the decryption result. 
Later on, the CSP disseminates the global sample size $N$ and sample sum $\bs$ to the local sites. 
Using these global information, the sample mean $\bar{\bx}$ can be easily computed as $\bs/N$. 

Similarly, each sites can get the desired standard deviation by the aid of the CSP. 
Specifically, the local sites encrypt the squared differences between their data and the global mean with divided by the global sample size $N$.
The server performs simple homomorphic additions of the encrypted variances from the sites. 
After that, the CSP decrypts the resulting ciphertext and each sites can get the desired global variance. 
In the end, the local data is mean-centered by subtracting the global mean from each column and divided by the standard deviation.

\begin{algorithm}[!htb]
\caption{Precomputation Phase - Standardization}
\label{alg:pre1}
\begin{algorithmic}
\State \textbf{[At local sites]:}
\For{$k=1$ {\bfseries to} $K$}
\State  $\bar{\bx}_k \la \frac{1}{n_k} \sum_{i=1}^{n_k} \bx_i^k$  \qquad \quad \qquad \qquad $\vartriangleright$ local mean
\State  Send $\Enc(n_k)$ and $\Enc(\bar{\bx}_k)$ to the server
\EndFor
\vspace{1.7mm}
\State  \textbf{[At the cloud server]:}  
\State   $\ct_{\mathsf{sample}}\leftarrow \sum_k \Enc(n_k)$
\State   $\ct_{\mathsf{sum}}\leftarrow \sum_k (\Enc(n_k)\cdot \Enc(\bar{\bx}_k))$ 
\State  Send $\ct_{\mathsf{sample}}$ and $\ct_{\mathsf{sum}}$ to the CSP
\vspace{1.7mm}
\State  \textbf{[At the CSP]:} 
\State  $N \leftarrow \rd{\Dec(\ct_{\mathsf{sample}})}$ \qquad \qquad $\vartriangleright$ global sample size
\State  $\bs \leftarrow \Dec(\ct_{\mathsf{sum}})$	\qquad \qquad \qquad $\vartriangleright$ scaled global mean
\State  Disseminate $N$ and $\bs$ to each local site
\vspace{1.7mm}
\State  \textbf{[At local sites]:}
\For{$k=1$ {\bfseries to} $K$}
\State  $\bar{\bx} \leftarrow \bs / N$ 	\qquad \quad \qquad \quad \qquad \qquad$\vartriangleright$ global mean
\State  $\bS_k \la \frac{1}{N} \sum_{i=1}^{n_k} (\bx_i^k-\bar{\bx})^2$
\State  Send $\Enc(\bS_k)$ to the server
\EndFor
\vspace{1.7mm}
\State  \textbf{[At the cloud server]:}  
\State   $\ct_{\mathsf{var}}\leftarrow \sum_k \Enc(\bS_k)$
\State  Send $\ct_{\mathsf{var}}$ to the CSP
\vspace{1.7mm}
\State  \textbf{[At the CSP]:} 
\State $\bS \leftarrow \Dec(\ct_{\mathsf{var}})$	\qquad \qquad \qquad \qquad $\vartriangleright$ global variance
\State  Disseminate $\bS$ to each local site
\vspace{1.7mm}
\State  \textbf{[At local sites]:}
\For{$k=1$ {\bfseries to} $K$}
\For{$i=1$ {\bfseries to} $n_k$}
\State  $\bx_{i}^{k} \leftarrow (\bx_{i}^{k} - \bar{\bx})/\sqrt{\bS}$
\EndFor
\EndFor
\end{algorithmic}
\end{algorithm}

\subsection{{Normalization}}\label{append:pre2}

Algorithm~\ref{alg:pre2} provides an explicit description of our secure protocol for data normalization. In this protocol, the data is securely divided by a quantity proportional to the maximum between its $\ell_2$ norm. 
This protocol is similar to~\cite{KLOJ19}, but is generalized to compute the maximum between multiple encrypted values. 

After the standardization process, each local site encrypts the maximum between the $\ell_2$ norm of their local data and transmits it to the server.
To be precise, the $k$-th local maximum $\mathsf{m}_k$ can be computed by $\mathsf{m}_k=\mmax\{||\bx_{i}^{k}||_2\}_{1 \le i \le n_k}$.
In the following, the server and CPS engage in computation to obtain an encryption of the global maximum. 
The server first generates two positive random numbers $r_1 \leftarrow \bbZ_{t_1}, r_2 \leftarrow \bbZ_{t_2}$ for some $t_1,t_2$ (which will be determined later), and obtains $K$ many ciphertexts by computing 
$\ct_k' = (\ct_k+r_1) \cdot r_2,$
where $\ct_k$ is a ciphertext of  $\mathsf{m}_k$ scaled by a factor of $\Delta$. 
Next, the server re-arranges the output ciphertexts and sends them to the CSP, so that the CSP does not know which local sites has the maximum. 
Suppose that $i_1,i_2,\ldots,i_K$ are distinct elements of $\{1,2,\ldots,K\}$.
Then each ciphertext $\ct_{i_k}$ is decrypted as the value $\sfM_{i_k}$, which is approximate to $(\Delta \sfm_{i_k} + r_1)\cdot r_2$. 
If assuming that $\Delta \sfm_{i_k}$ is less than $t_1$ and $t_1,t_2$ are co-prime, any element $\sfM_{i_k}$ is statistically close to uniform in $\bbZ_{t_1\cdot t_2}$. 
Hence, the CSP learns nothing else after decryption while computing the randomized maximum value between them, say $\sfM = \mmax\{\sfM_{i_k}\}_{1 \le k \le K}$.
Later on, an encryption of this value $\sfM$ is sent to the server.
Afterwards, the server multiplies this ciphertext by $r_2^{-1}$ and subtracts $r_1$, yielding a ciphertext which represents the desired constant $\sfm=\mmax\{||\bx_{i}^{k}||_2\}_{1 \le k < K,1 \le i \le n_k}$ since
\begin{align*}
    r_2^{-1} \cdot \Enc(\sfM) - r_1 & \approx \Enc\left(r_2^{-1} \sfM - r_1\right)  \approx \Enc\left(r_2^{-1}\cdot(\Delta \sfm + r_1)\cdot r_2 - r_1\right)  \approx \Enc(\Delta \sfm). 
\end{align*}
Finally, the server sends the output ciphertext to the CSP and its decrypted result is disseminated to each site. 

\begin{algorithm}[!htb]
\caption{Precomputation Phase - Normalization}
\label{alg:pre2}
\begin{algorithmic}
   \State \textbf{[At local sites]:}
   \For{$k=1$ {\bfseries to} $K$}
   \State  $\ct_k \leftarrow \Enc\left(\mmax\{||\bx_{i}^{k}||_2\}_{1 \le i \le n_k}\right)$ \qquad $\vartriangleright$ local max
   \State  Send $\ct_k$ to the server
   \EndFor
   \vspace{2mm}
    \State  \textbf{[At the cloud server]:} 
    \State  Generate two random numbers $r_1, r_2$
    \For{$k=1$ {\bfseries to} $K$}
    \State  Compute $\ct'_k \leftarrow (\ct_k + r_1)\cdot r_2$
    \EndFor
    \State  Send $\{\ct'_{i_k}\}_{1 \le k \le K}$ to the CSP
    \vspace{2mm}
    \State  \textbf{[At the CSP]:} 
    \State  Compute $\sfM \leftarrow \mathsf{max}\{\Dec(\ct'_{i_k})\}_{1 \le k \le K}$ \quad  
    \State  Send $\Enc(\sfM)$ to the server \qquad 
   \vspace{2mm}
    \State  \textbf{[At the cloud server]:} 
    \State Compute $\ct_{\mmax} \leftarrow r_2^{-1}\cdot \Enc(\sfM) - r_1$
    \State Send $\ct_{\mmax}$ to the CSP
    \vspace{2mm}
    \State  \textbf{[At the CSP]:} 
    \State  Compute $\sfm \leftarrow \Dec(\ct_{\mmax})$ \qquad \qquad \quad $\vartriangleright$ global max
    \State  Disseminate $\sfm$ to each local sites 
    \vspace{2mm}
   \State  \textbf{[At local sites]:} 
   \State  Compute $\bx_{i}^{k} \leftarrow \bx_{i}^{k}/\sfm$
\end{algorithmic}
\end{algorithm}

\newcommand{\R}{{\mathcal R}}
\section{HE Parameter Setting}~\label{append:hepar}

Our underlying HE scheme is based on the Ring Learning with Error (RLWE) assumption over the cyclotomic ring $\R = \bbZ[x]/(x^n+1)$ for a power-of-two integer $n$.
Let us denote by $[\cdot]_q$ the reduction modulo $q$ into the interval $(-q/2,q/2]\cap \bbZ$ of the integer. 
We write $\R_q = \R/q\R$ for the residue ring of $\R$ modulo an integer $q$. 
As mentioned in Section 2.2, the CKKS scheme adapts some Discrete Fourier Transformation (DFT)-like algorithms
to transform an $(n/2)$-dimensional complex vector into an element of the cyclotomic ring.
So, a plaintext vector can be encrypted into a single packed ciphertext and one can perform operation on these plaintext values in parallel. 

For the sake of optimization of the basic polynomial arithmetic, we use the Residue Number System (RNS) variant of CKKS~\cite{CHK+18b}. 
If taking a ciphertext modulus $Q=\prod_{i=0}^\ell q_i$ which is a product of distinct primes, a polynomial with a large modulus $Q$ can be represented as a tuple of polynomials with smaller coefficients modulo $q_i$.
If needed, we raise a ciphertext modulus from $Q_j=\prod_{i=0}^j q_i$ to $PQ_j$ for a prime number $P$, called the \textit{special modulus}, and perform the key-switching procedure over $R_{PQ_j}$ followed by modulus reduction back to $Q_j$. 
This series of process is usually done after homomorphic multiplication or rotation operation. 
We note that the RNS primes should be 1 modulo $2n$ to utilize an efficient Number Theoretic Transformation (NTT) algorithm.

We begin with the parameter $\ell$ which determines the largest bit-size of a fresh ciphertext modulus. 
As discussed in Section 4.1, we need to set the number of levels $\ell$ to be at least $\ell \ge 6$ for the evaluation of our protocol. 
We note that $q_0$ is the output ciphertext modulus and the final resulting ciphertext at iteration $t$ represents the desired model parameters $\btheta_t$ but is scaled by a factor of $\Delta$. This implies that $q_0$ should be larger than $\Delta$ to ensure correctness of decryption. 
We set the scale factor $\Delta=2^{45}$ in the light of the encoding/encryption errors of size $O(n)$,
and thus we use the RNS primes of sizes roughly $\log P \approx 55$, $\log q_0 \approx 55$, and $\log q_i \approx 45$ for $1\le i \le \ell$. 
Therefore, we derive a lower bound on the bit size of the largest RLWE modulus as 
\begin{align*}
 \log PQ &= \log P + \log q_0 + \ell  \log q_i \approx 55 + 55 + 6 \cdot 45 = 380.
\end{align*}
We set the secret distribution as the uniform distribution over the set of polynomials whose coefficients are in $\{0,\pm 1\}$.
Each coefficient of an error is drawn according to the discrete Gaussian distribution centered at zero with standard deviation $\sigma=3.2$.
We followed the recommended parameters from the homomorphic encryption standardization (HES) workshop paper~\cite{HESecurityStandard}, and so we took the ring dimension $n= 2^{14}$ to provide at least 128-bits security level of our parameters.
On the other hand, the parameters $t_1$ and $t_2$ are chosen to be 50-bit and 25-bit primes, respectively.

\section{Experimental Settings and Results}\label{append:results}

\subsection{Detailed information of the auxiliary constant $\boldsymbol{c}$}
The values of the auxiliary constant $c$ that we used in each experiment are described in Table~\ref{tbl:c}.

\begin{table}[h]
\caption{Auxiliary constant $c$ for different experimental settings.}\label{tbl:c}
\renewcommand{\arraystretch}{1.21}  
\renewcommand{\tabcolsep}{1.9mm}   
\vskip 0.05in
\begin{center}
\begin{tabular}{cccccc}
\toprule
\multirow{2}{*}{\textbf{{Task}}}& \multicolumn{5}{c}{\textbf{{The number of sites}}} \\
 & 1 & 2 & 5 & 10 & 16 \\
\midrule
{Density estimation} & {80} & {80} & {-} & {-} &  {-} \\
\midrule
{Logistic regression} & {3500} & {3500} & {1000} & {1000} &  {900} \\
\midrule
{Survival analysis} & {2000} & {2000} & {800} & {-} &  {-} \\
\bottomrule
\end{tabular}
\end{center}
\vskip -0.1in
\end{table}

\subsection{Detailed descriptions on the experiments of Bayesian logistic regression and Bayesian survival analysis}\label{append:description}
\begin{itemize}
    \item  [$\bullet$]  
We applied our approach to Bayesian logistic regression model. Logistic regression is very popular in the field of health informatics, where sensitive data are distributed across institutions (\ie hospitals). 
The logistic function of the $i$-th output $y_i \in \{-1,+1\}$ given the corresponding input vector $\bx_i\in \mathbb{R}^d $ can be modelled as follows: $\Pr(y_i|\bx_i,\bbeta) =  {1}/{(1+\exp(-y_i\bbeta^T \bx_i))}$,
where $\bbeta$ are the parameters.
Given the log likelihood of logistic regression
$$ l(\bbeta)= - \sum_{i=1}^{N}\log(1+\exp(-y_i\bbeta^T \bx_i)),$$
the gradient at time $t$ in a distributed setting is as follows:
$$ l'(\bbeta_{t}) =  \sum_{k=1}^{K} (\bX^k)^T(\boldsymbol{y}^k-\boldsymbol{\mu}_t^{k}),$$
where $\bmu_{t}$ is an $N$-dimensional vector with the $i$-th element $\Pr(Y=1|\bx_i,\bbeta_{t})$. 

The PhysioNet Challenge 2012 dataset used in this experiment is extracted from the Multiparameter Intelligent Monitoring in Intensive Care \RNum{2} database comprised of patient stays in the intensive care unit (ICU) lasting at least 48 hours to predict mortality. We used the dataset A consisting of 4,000 subjects whose age at ICU admission was 16 years or over. The data were formatted as time-stamped measurements for 37 distinct variables and four static variables. We transformed each time-series variable into min, max, mean, first value, and last value variables as a way to summarize it. Missing values were replaced by the mean value of a variable and the columns having all the same values were removed. 
    \item  [$\bullet$]  Survival analysis is a class of statistical methods to study the time until the occurrence of a specified event. The usual methods assume that all individuals under study are subjects to the event the interest. Among several models for survival analysis, we focus on Exponential distribution-based parametric survival model. 
Given $\{\bx_i,c_i,t_i\}_{i=1}^{N}$, where $c_i\in \{ 0,1 \}$ is a censoring indicator whether the $i$-th sample is censored (0) or not (1) and $t_i$ is the observed response for the $i$-th sample (\ie $t_i = \min(C_i,T_i)$ when $C_i$ and $T_i$ denote time to the specified event and time to be censored respectively),
the log likelihood function of  Exponential model with parameters $\alpha, \bbeta $ is
$$ l(\alpha,\bbeta)= - \sum_{i=1}^{N} \left(c_i(\log \alpha+\bbeta^T\bx_i) + \frac{t_i}{\alpha\exp(\bbeta^T\bx_i)}\right), $$
and the gradients of $\alpha, \bbeta$ at iteration $t$ are  
\begin{align*}
l'(\alpha_{t}) &=  \sum_{k=1}^{K} \frac{(\boldsymbol{t}^k)^T \textbf{1}}{\exp{(\bbeta_t^T\bX^k \textbf{1})}\cdot(\boldsymbol{c}^k)^T \textbf{1}},\\
l'(\bbeta_{t}) &= \frac{1}{\alpha_t} \sum_{k=1}^{K} \frac{(\bX^k)^T\boldsymbol{t}^k}{\exp{(\bbeta_t^T\bX^k \textbf{1})}}-\sum_{k=1}^{K}  (\bX^k)^T\textbf{1},
\end{align*}
respectively. We conducted the experiment with $\alpha=1$ for simplicity. 
Original FLchain dataset has 8 variables, but we deleted one of them, the starting year, which is actually an unnecessary variable. 

\end{itemize}

\begin{figure}[t]
\begin{center}
\hspace*{\fill}
\subfloat[{\small{Learning curves}}]{\label{appfig:logreg_lraucbox}
\includegraphics[width=0.49\columnwidth]{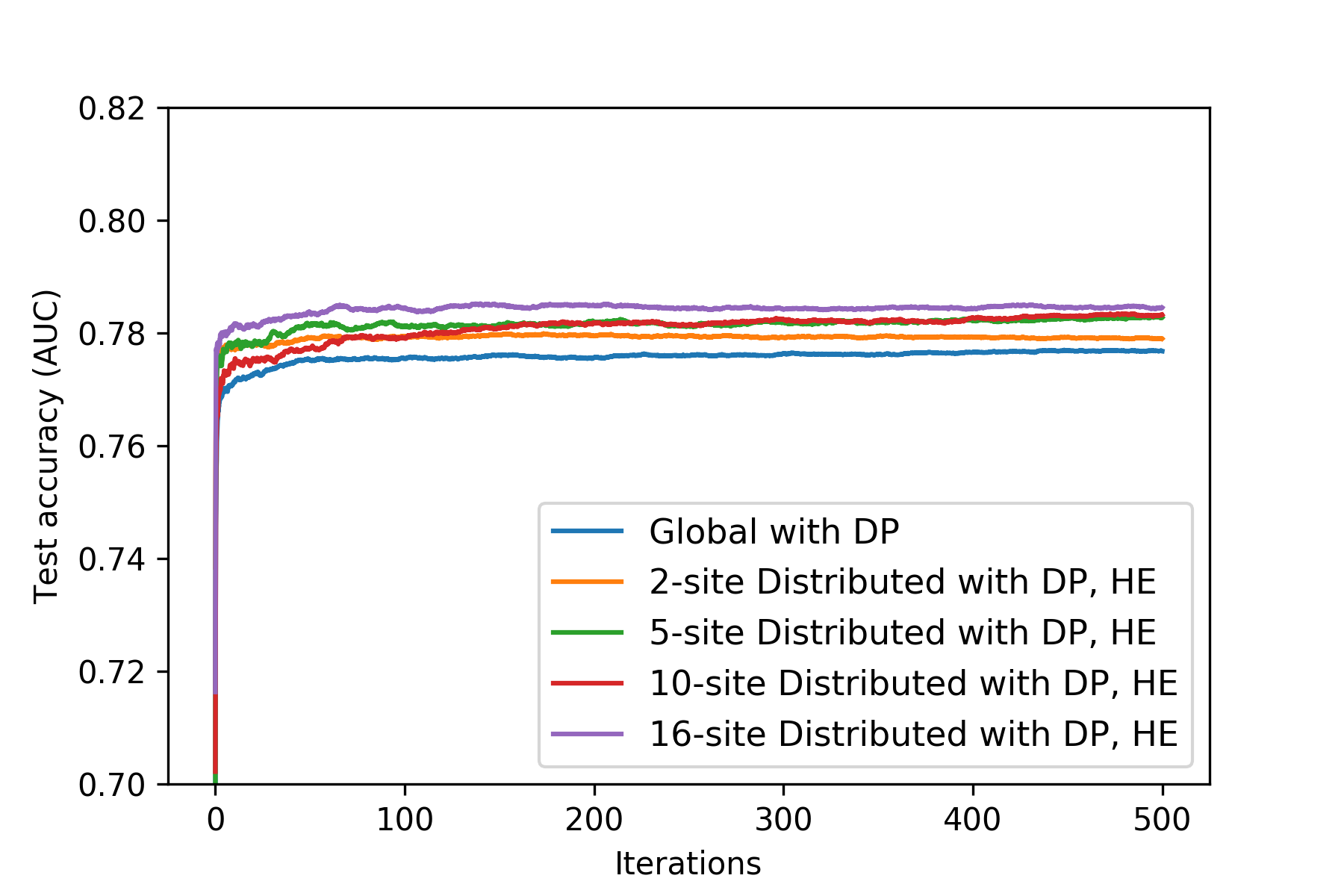}} \hfill
\subfloat[{\small{Classification accuracies}}]{\label{appfig:logreg_accuracy}
\includegraphics[width=0.49\columnwidth]{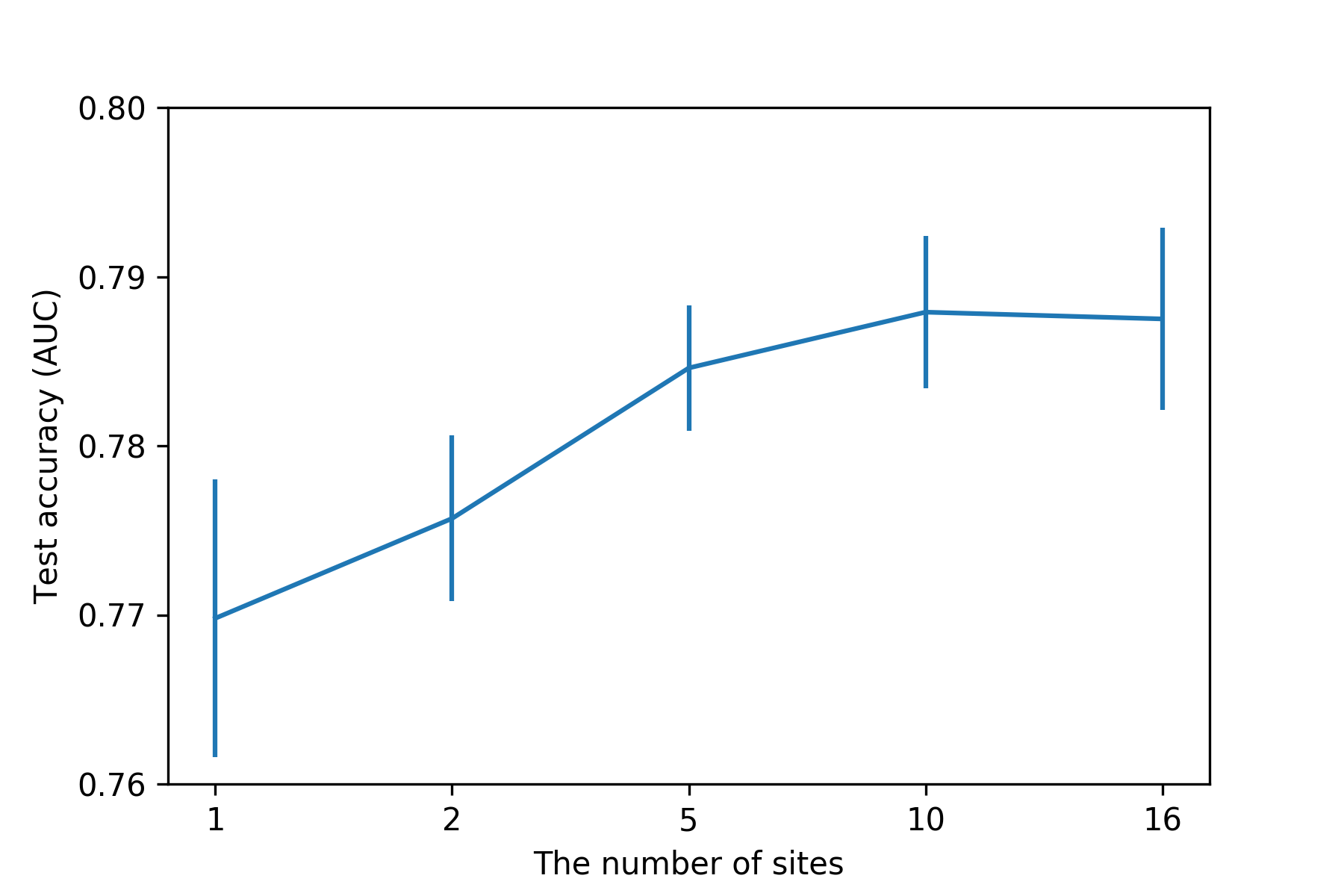}} \hfill
\hspace*{\fill}
\caption{Bayesian logistic regression on the PhysioNet dataset at various numbers of sites, when $\epsilon=1$.}
\label{appfig:logreg}
\end{center}
\end{figure}

\vspace{-6mm}

\begin{figure}[t]
\begin{center}
\includegraphics[width=0.55\columnwidth]{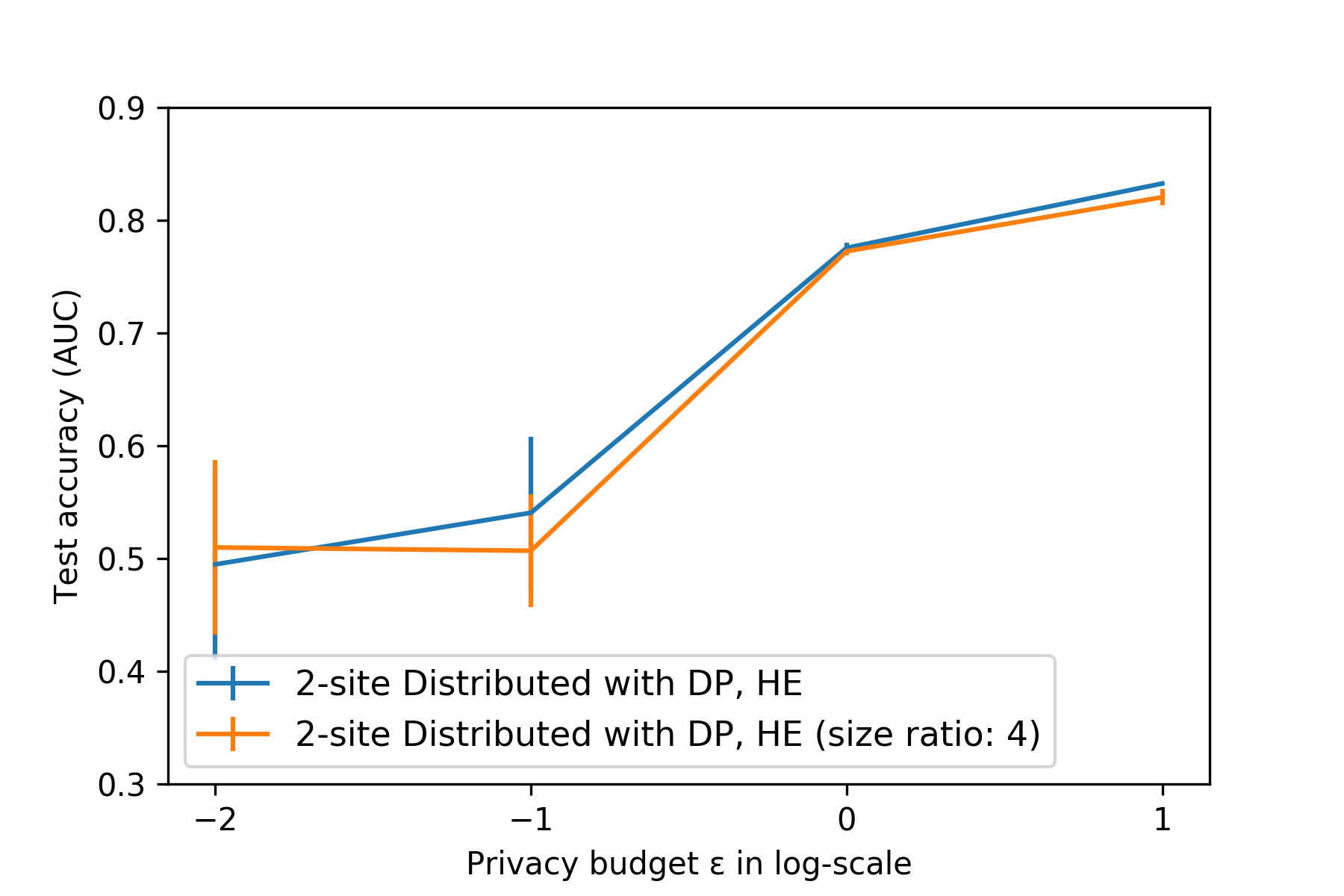}
\caption{Classification accuracies of Bayesian logistic regression with the different local sample sizes.}
\label{appfig:imbalanced}
\end{center}
\end{figure}

\subsection{Experimental results at the various numbers of local sites $\boldsymbol{K}$}\label{append:results1}

Additional experimental results are described when differentiating the number of local sites from 2 to 16 with the privacy budget $\epsilon=1$. Here, we only conducted experiments on Bayesian logistic regression since we have already seen that the tendency of prediction results between Bayesian logistic regression and Bayesian survival analysis are the same. As shown in Figure~\ref{appfig:logreg}, our approach allows us to achieve robust prediction accuracy with no significant difference from the global model even if the number of sites increases to 16, which is quite useful. The differences of results among the various numbers of sites might be caused by the different choice of auxiliary constant, and the different sample spaces; since we fixed the batch size as 320, the sample space per each site would be reduced when the number of sites increases. 


\subsection{Experimental results with different local sample sizes}\label{append:results2}
To check the prediction performance of our approach when local parties have different sample sizes, another simple experiment on 2-site distributed Bayesian logistic regression was performed with the privacy budget from -2 to 1 in the log-scale. We divided the PhysioNet dataset into two with a sample size ratio of four to one to simulate a scenario with the large-sized local site and small-sized local site; the target (\ie label) ratio and batch size remain the same with the basic experimental setting. Figure~\ref{appfig:imbalanced} shows that our approach is not sensitive to the local sample size, which means that our approach can be practical even in the situation with participating parties having different local sizes, for example, a large hospital and a small hospital. 

\end{document}